\documentclass[11pt,aps,prd,preprint,showpacs,groupedaddress,nofootinbib]{revtex4}
\usepackage{amsmath,amssymb,amsbsy,color}

\usepackage{psfrag}  

\newcommand{\non}{\nonumber}
\def\dfrac#1#2{\displaystyle\frac{#1}{#2}}

\newcommand{\pslash}{p\kern-1ex /}
\newcommand{\lslash}{l\kern-1ex /}
\newcommand{\sslash}{s\kern-1ex /}
\newcommand{\Dslash}{{\cal D}\kern-1.5ex /}

\newcommand{\tr}{{\rm tr}}

\newcommand{\beqa}{\begin{eqnarray}}
\newcommand{\eeqa}{\end{eqnarray}}
\newcommand{\be}{\begin{equation}}
\newcommand{\ee}{\end{equation}}
\newcommand{\bea}{\begin{eqnarray}}
\newcommand{\eea}{\end{eqnarray}}
\newcommand{\bay}{\begin{array}}
\newcommand{\eay}{\end{array}}

\newcommand{\pref}[1]{(\ref{#1})}
\def\chpt{\raise0.4ex\hbox{$\chi$}PT}
\newcommand{\pal}{\partial}
\newcommand{\ba}{\begin{eqnarray}}
\newcommand{\ea}{\end{eqnarray}}

\newcommand{\shouleft}
           {\left ( \mbox{\hspace{-1mm}} \frac{}{} \mbox{\hspace{-1mm}} \right.}
\newcommand{\shouright}
           {\left. \mbox{\hspace{-1mm}} \frac{}{} \mbox{\hspace{-1mm}} \right)}

\newcommand{\daileft}
           {\mbox{\hspace{-1mm}}\left [ \mbox{\hspace{-1mm}} \frac{}{} \mbox{\hspace{-1mm}} \right.}
\newcommand{\dairight}
           {\left. \mbox{\hspace{-1mm}} \frac{}{} \mbox{\hspace{-1mm}} \right]}

\begin{document}

%\draft
\preprint{UTHEP-511}
\preprint{UTCCS-P-17}

\title{
Vector meson masses in 2+1 flavor Wilson Chiral Perturbation Theory
}
\author{S.~Aoki}
\affiliation{Graduate School of Pure and Applied Sciences, University  
of Tsukuba, Tsukuba, Ibaraki  305-8571, Japan}
\affiliation{
Riken BNL Research Center, Brookhaven National Laboratory, Upton, New York 11973,
USA}
\author{O.~B\"ar}\thanks{Present address: 
Institut f\"ur Physik,
Humboldt Universit\"at zu Berlin,
Newtonstr.~15, 12489 Berlin, Germany.}
\affiliation{Graduate School of Pure and Applied Sciences, University  
of Tsukuba, Tsukuba, Ibaraki  305-8571, Japan}
\author{S.~Takeda$^{\ast}$}
\affiliation{Graduate School of Pure and Applied Sciences, University  
of Tsukuba, Tsukuba, Ibaraki  305-8571, Japan}

\date{\today}
%
%===========
\begin{abstract}
%===========
%
We calculate the vector meson masses  
in $N_{\rm f} = 2+1$ Wilson chiral perturbation theory
at next-to-leading order.
Generalizing the framework of heavy vector meson chiral perturbation theory,
the quark mass and the lattice cutoff  dependence
of the vector meson masses is derived.
Our chiral order counting 
%for the Wilson chiral perturbation theory
assumes that the lattice cut-off artifacts are of the order
of the typical pion momenta, $p \sim a\Lambda_{\rm QCD}^{2}$.  
This counting scheme is 
consistent with the one in the pseudo scalar meson sector
where the O($a^2$) terms are included in the leading order chiral Lagrangian.
\end{abstract}

\maketitle
%\narrowtext

%1
%=================
\section{Introduction}
%=================
%

This is the second in a series of papers where we compute a variety of
mesonic quantities in 2+1 flavor Wilson Chiral Perturbation Theory
(Wilson \chpt). After having computed the pseudo scalar mesons masses
\cite{Aoki:20051} we present here the results for the vector meson
masses. The calculation of the pseudo scalar decay constants and the
axial vector Ward identity quark mass is in progress
\cite{Aoki:20053}. The main goal of this series of papers is to provide
the necessary chiral fit forms for unquenched 2+1 flavor Lattice QCD
simulations with improved Wilson fermions, as they have been currently
performed by the CP-PACS/JLQCD collaboration \cite{Ishikawa:2004xq}. 

The quark masses in the CP-PACS/JLQCD simulations are heavier than their
physical values. The ratio of the pseudo scalar to vector meson mass is
in the range $m_{PS}/m_{V}\simeq 0.62 - 0.78$.  A chiral extrapolation
in the light up and down quark masses is thus required. In order to
perform the chiral extrapolation before taking the continuum limit we
formulate \chpt\ at non-zero lattice spacing, as originally proposed in
Refs.\ \cite{Sharpe:1998xm,Lee:1999zx}. A variety of pseudo scalar
quantities has been already computed, mainly for 2 flavor Wilson \chpt\
(see Ref.\ \cite{Bar:2004xp} and references therein). Although the
vector meson masses were calculated recently in 2 flavor partially
quenched Wilson \chpt\ \cite{Grigoryan:2005zj}, 
the result for 2+1 flavors was missing.

In this paper we follow the heavy vector meson formalism first
introduced by Jenkins {\em et.\ al.\ }in Ref.\
\cite{Jenkins:1995vb}. The generalization to Lattice QCD at non-zero
lattice spacing $a$ is straightforward and mirrors the strategy spelled
out in Refs.\ \cite{Sharpe:1998xm,Lee:1999zx}. However, since the
lattice spacing is an additional expansion parameter the power counting
requires some care. Here we adopt a power counting which assumes that
lattice cut-off artifacts are of the order of the typical pion momenta,
$ a\Lambda_{\rm QCD}^{2} \sim p$, as it was assumed in Ref.\ \cite{Aoki:2003yv}.
Previous results in unquenched
2-flavor simulations \cite{Namekawa:2004bi} seem  to indicate that this
is the appropriate power counting for describing the lattice results of
the CP-PACS/JLQCD collaboration. 

There is a price to pay if one wants to perform the chiral extrapolation
before taking the continuum limit. The chiral fit forms contain terms
proportional to powers of the lattice spacing accompanied by additional
unknown low-energy constants. These constants are essentially
unconstrained and serve as additional fit parameters. Obviously, the presence of too
many of these additional parameters would limit or even spoil the chiral
extrapolation. 
Our one loop results for the $\rho$ and $K^{\ast}$ meson masses contain
seven unknown fit parameters compared to three in the corresponding continuum \chpt\ result
of Ref.\ \cite{Jenkins:1995vb}. This number is still small enough for
the results to be useful for the chiral extrapolation of the
CP-PACS/JLQCD collaboration data.

This paper is organized as follows.
In Sec.\ \ref{sec:VMChPT}, 
we briefly review the heavy meson formalism, as it was introduced
in Ref.\ \cite{Jenkins:1995vb}.
In Sec. \ref{sec:VMWChPT} we first summarize the meson and spurion
fields which are necessary in our calculation. After discussing the
power counting we derive the chiral effective Lagrangian and compute the
vector meson masses to one loop. Concluding remarks are given in section
\ref{sec:Conclusion}.

%2
%=================================
\section{Heavy vector meson effective theory}
\label{sec:VMChPT}
%=================================
%
%As mentioned in the introduction,
We adopt the heavy vector meson effective theory
to derive the quark mass and lattice spacing dependence of the vector meson 
masses.\footnote{This formalism is very similar to heavy baryon chiral
perturbation theory \cite{Jenkins:1990jv}.}
This effective theory can deal with vector number conserving
decay processes like 
$V \longrightarrow V^{\prime} X$, where
$V$ and $V^{\prime}$ are vector mesons, and 
$X$ represents a state with one or more low momentum pseudo scalar mesons.
One the other hand, it cannot be applied to the process
$\rho \longrightarrow \pi \pi$, for example, 
since this decay includes hard pions in the final state.

Some physical quantities have been calculated
within the continuum formulation of the heavy vector meson effective theory.
The first result for the vector meson masses 
up to  O($p^3$) can be found in Ref.\ \cite{Jenkins:1995vb}.
The effects of isospin breaking and
electromagnetic corrections 
are included in Ref.\ \cite{Bijnens:1996kg}.
The calculation of the vector meson masses to O($p^4$)
was done in Ref.\ \cite{Bijnens:1998di},
while the results for the decay constants
are given in Ref.\ \cite{Bijnens:1997ni}.
For results in quenched and partially quenched
\chpt,  see
Ref.\ \cite{Booth:1996hk} and
Ref.\ \cite{Chow:1997dw}, respectively.

Let us briefly review the main idea behind the heavy vector meson formalism.
The following argument can be found in Ref.\ \cite{Bijnens:1997ni}.
The vector meson sector introduces
a typical energy scale, the vector meson mass $m_V$.
This mass does not vanish in the chiral limit, and
it is of about the same size as the chiral symmetry breaking scale,
$4\pi f_{\pi} \sim 1$ GeV.
The vector mesons are described by heavy matter fields.
In order to
obtain the Lagrangian of
the effective theory one expands
the relativistic Lagrangian 
in powers of $1/m_V$.
The starting point is the free relativistic vector meson Lagrangian,
\be
  {\cal L}_R
  =
  -\frac{1}{4} V^{\mu\nu} V_{\mu\nu}
  +\frac{1}{2} m_V^2 V^{\mu} V_{\mu},
\ee
with $V_{\mu\nu}=\pal_{\mu} V_{\nu} - \pal_{\nu} V_{\mu}$.
The relativistic field $V_{\mu}$ is decomposed into two parts,
a parallel and a perpendicular component with respect to the
4-velocity $v_{\mu}$ ($v^2=1$) of the vector meson,
\be
  V_{\mu}
  =
  P_{\mu\nu} V^{\nu} + v_{\mu} (v \cdot V)
  =
  V_{\perp\mu} + v_{\mu} V_{\parallel},
\ee
where
\bea
P_{\mu\nu}& =& g_{\mu \nu} - v_{\mu} v_{\nu},
\eea
is the projector which selects the perpendicular
component of the vector field (
$P^2=P$, $v^{\mu} P_{\mu \nu} =0$).
$V_{\perp\mu}$ and $V_{\parallel}$
can further be written as
\ba
  V_{\perp\mu}
  &=&
  \frac{1}{\sqrt{2 m_V}}
  [e^{-i m_V v \cdot x} W_{\mu}
  +e^{i m_V v \cdot x} W_{\mu}^{\dag}], 
\\
  V_{\parallel}
  &=&
  \frac{1}{\sqrt{2 m_V}}
  [e^{-i m_V v \cdot x} W_{\parallel}
  +e^{i m_V v \cdot x} W_{\parallel}^{\dag}], 
\ea
where $W_{\mu}$ and $W_{\mu}^{\dag}$
are effective vector meson fields,
and $P_{\mu\nu} W^{\nu}= W_{\mu}$ and
$P_{\mu\nu} W^{\nu\dag}= W_{\mu}^{\dag}$
are understood.
By neglecting terms proportional to
$\exp{(\pm 2 i m_V v \cdot x)}$, which
oscillate rapidly if one takes $m_V$ to infinity,
the Lagrangian ${\cal L}_R$ in terms of the $W$ fields is given by
\ba
  {\cal L}_R
  &\longrightarrow&
  \frac{1}{2 m_V}
  \left[ 
  - \pal_{\mu} W_{\parallel}^{\dag} \pal^{\mu} W_{\parallel}
  + (v \cdot \pal W_{\parallel}^{\dag})(v \cdot \pal W_{\parallel})
  \right]
+ \frac{m_V}{2} W_{\parallel}^{\dag} W_{\parallel}
\non \\
  && +
  \frac{1}{2}
  \left[
  - i \pal_{\mu} W_{\parallel}^{\dag} W^{\mu}
  + i W_{\mu}^{\dag} \pal^{\mu} W_{\parallel}
  \right]
\non \\
  &&+
  \frac{1}{2 m_V}
  \left[
    \pal_{\mu} W_{\parallel}^{\dag} (v \cdot \pal W^{\mu})
  + (v \cdot \pal W_{\mu}^{\dag}) \pal^{\mu} W_{\parallel}
  \right]
\non \\
  &&-
  i W_{\mu}^{\dag} ( v\cdot\pal) W^{\mu}.
 \label{eqn:lagrangianW}
\ea
In the infinite mass limit, $m_V \rightarrow \infty$,
the  $W_{\parallel}$ field decouples
because of the presence of the mass term 
$m_V W_{\parallel}^{\dag} W_{\parallel}/2$. 
In order to remove the parallel component, one can impose the constraint 
$v \cdot V = 0$ or $v \cdot W = 0$.
Alternatively, we can remove $W_{\parallel}$
by making use of the equation of motion,
\be
  W_{\parallel}
  =
  \frac{-i}{m_V} \pal_{\mu} W^{\mu}
  + O(1/m_V^2),
 \label{eqn:EOMWparallel}
\ee
which also shows that $W_{\parallel}$ is
suppressed by powers of $1/m_V$ relative
to the perpendicular component $W_{\mu}$.
Having removed the parallel component the resulting effective Lagrangian simplifies to 
\ba
  {\cal L}_R
&=&
  - i W_{\mu}^{\dag} ( v\cdot\pal) W^{\mu}+ O(1/m_V).
\ea
The first term is the kinetic term for the
vector meson in the heavy effective theory.
Vector meson fields which appear in the following
are the effective fields $W_{\mu}$ and $W_{\mu}^{\dagger}$.
The partial derivative $i\pal_{\mu}$ acting on 
these effective fields produces a small residual momentum $r_{\mu}$,
defined as
\be
  k_{\mu}
  =
  m_V v_{\mu} + r_{\mu},
\ee
where $k_{\mu}$ is the usual four-momentum of the vector meson.
We assume here that the residual momentum
is of the size of the low momentum of the pseudo scalar meson.
This assumption holds in interaction processes with soft pions.

%3
%=================================
\section{Wilson chiral perturbation theory for vector mesons}
\label{sec:VMWChPT}
%=================================
%
The chiral effective Lagrangian for heavy vector mesons is expanded in
powers of the small pseudo scalar momenta and masses, and the residual
momentum $r_{\mu}$ of the vector mesons. 
The symmetries of QCD, in particular chiral symmetry, dictate the form
of the terms in the chiral Lagrangian. Even though the quark masses
explicitly break chiral symmetry, their effect is properly taken into
account by a spurion analysis, where the quark mass matrix is assumed to
transform non-trivially under chiral transformations in an intermediate step. 

The same principles apply to Wilson \chpt, the low energy effective
theory for Lattice QCD with Wilson fermions. The main difference is the
presence of an additional expansion parameter, the lattice spacing $a$
\cite{Sharpe:1998xm,Rupak:2002sm}. The non-zero lattice spacing is also
taken into account  by the spurion analysis, and the transformation
behaviour of the corresponding spurion field is exactly the same as for
the quark masses. Constructing the terms in the chiral Lagrangian of
Wilson \chpt\ for vector mesons is therefore straightforward. Apart from
the presence of two instead of one spurion field there is essentially no
difference to Ref.\ \cite{Jenkins:1995vb}. The order counting, however,
is not entirely obvious, since both the quark masses and the lattice spacing are
expansion parameters and their relative size is important for a
consistent power counting, as we will discuss in section \ref{subsec:counting}.

%
%=================================
\subsection{Matter and spurion fields}
 \label{subsec:block}
%=================================
%
The pseudo scalar meson field is introduced as usual as an SU(3) unitary matrix
\be
  \Sigma
  = 
  \exp \left( \frac{2i\Pi}{f} \right),
\ee
where 
\be
  \Pi 
  =
  \sqrt{2} \pi^a T^a
  =
  \left[
    \begin{array}{ccc}
    \frac{\pi^0}{\sqrt{2}} + \frac{\eta}{\sqrt{6}}
&   \pi^{+}
&   K^{+}
\\
    \pi^{-}
&  -\frac{\pi^0}{\sqrt{2}} + \frac{\eta}{\sqrt{6}}
&   K^{0}
\\
    K^{-}
&   \bar{K}^{0}
&  -\frac{2\eta}{\sqrt{6}}
    \end{array}
  \right].
\ee
The SU(3) generators $T^a$ ($a=1,\ldots,8$) are normalized such that
\be
\mbox{tr}[T^a T^b]  = \frac{1}{2} \delta^{ab},
\ee
and $f$ is the leading order pseudo scalar decay constant in the chiral
limit.\footnote{Our normalization corresponds to $f\approx132$ MeV. }
The field $\Sigma$ transforms according to $\Sigma \rightarrow L\Sigma
R^{-1}$ under chiral rotations with  $L \in SU(3)_{L}$ and $R\in
SU(3)_{R}$. The square root of $\Sigma$,
\bea
\xi & =& \exp \left( \frac{i\Pi}{f} \right) \,=\, \sqrt{\Sigma},
\eea
is needed to describe the interaction of the pseudo scalars with the
vector mesons. It transforms under chiral transformations according to 
\bea\label{DefU}
\xi &\rightarrow & L\xi U^{\dagger}\,=\,U\xi R^{-1},
\eea
with an SU$(3)$ matrix  $U$. In fact, eq.\ \pref{DefU} defines $U$,
which is a function of $L,R,$ and $\Pi$. For vector transformations with
$L=R$ one finds $U=L=R$. 

The vector meson fields are introduced as an octet
\be
  {\cal O}_{\mu}
  =
  \sqrt{2} \rho^{a}_{\mu} T^a
  =
  \left[
    \begin{array}{ccc}
    \frac{\rho^0_{\mu}}{\sqrt{2}} + \frac{\phi^{(8)}_{\mu}}{\sqrt{6}}
&   \rho^{+}_{\mu}
&   K^{\ast +}_{\mu}
\\
    \rho^{-}_{\mu}
&  -\frac{\rho^0_{\mu}}{\sqrt{2}} + \frac{\phi^{(8)}_{\mu}}{\sqrt{6}}
&   K^{\ast 0}
\\
    K^{\ast -}_{\mu}
&   \bar{K}^{\ast 0}_{\mu}
&  -\frac{2\phi^{(8)}_{\mu}}{\sqrt{6}}
    \end{array}
  \right],
\ee
and a singlet  $S_{\mu}=\phi^{0}_{\mu}$.
These fields are required to satisfy the constraints
\be
  v \cdot S =  v \cdot {\cal O} = 0,
\ee
in order to describe spin 1 particles with three polarization states. As
stated in the previous section, this constraint can be enforced by
applying the projector $P_{\mu \nu}$ on the vector meson fields. In the
following we assume that this projector has been applied, i.e.\ we
implicitly assume
$S_{\mu}=P_{\mu \nu}S^{\nu}$ and ${\cal O}_{\mu}= P_{\mu \nu}{\cal O}^{\nu}$.

For the construction of the chiral Lagrangian the quark masses and the lattice spacing are treated as spurion fields. For the quark mass matrix we use
\ba
  \tilde{M}_q
  &=&
  \mbox{diag} ( \tilde{m},  \tilde{m}, \tilde{m}_s )
  =
  \tilde{M}_0 I + \tilde{M}_8 T^8,
\ea
where $\tilde{M}_0$ and $\tilde{M}_8$ are expressed in terms of
$\tilde{m}$ and $\tilde{m}_s$
\be
  \tilde{M}_0
  =
  \frac{2\tilde{m}+\tilde{m}_s}{3},
\mbox{\hspace{5mm}}
  \tilde{M}_8
  =
  \frac{2(\tilde{m}-\tilde{m}_s)}{\sqrt{3}}.
\ee
We use the tilde in order to highlight that 
the quark masses $\tilde{m}$ and $\tilde{m}_s$ denote shifted
quark masses, which are defined such that the tree level pseudo scalar meson masses
become zero if $\tilde{m}=\tilde{m}_s=0$. See Ref.\ \cite{Aoki:20051} and also
appendix \ref{sec:feynman} for details.

We introduce a spurion field $A$ to include the effect of a non-zero
lattice spacing $a$. This field transforms under chiral transformations
just as the quark mass field $M$ ($M=\tilde{M}_q$ after the spurion
analysis), i.e.
\bea \label{eqn:spurionA}
  A  & \rightarrow & LAR^{-1}.
\eea
Once the chiral Lagrangian is derived, the spurion is set to $a I$ where
$I$ denotes the unit matrix in flavor space.

For the construction of the chiral Lagrangian it will also be useful to
introduce the following quantities:
\ba
M_{\pm} 
&=&
\frac{1}{2}(\xi M^{\dag} \xi \pm \xi^{\dag} M \xi^{\dag}),
\\
W_{\pm}
& =&
\frac{1}{2}(\xi A^{\dag} \xi \pm \xi^{\dag} A \xi^{\dag}),
\\
V_{\mu}
&=&
  \frac{1}{2}(\xi \partial_{\mu} \xi^{\dag} 
            + \xi^{\dag} \partial_{\mu} \xi),\\
A_{\mu}
&  = &
\frac{i}{2}(\xi \partial_{\mu} \xi^{\dag} - \xi^{\dag} \partial_{\mu} \xi),
\label{DefAmu}\\  
{\cal D}_{\nu}{\cal O}_{\mu}
&=&
\partial_{\nu}{\cal O}_{\mu} +[V_{\nu},{\cal O}_{\mu}].
\ea
The transformation behavior of these quantities as well as of the meson and
spurion fields under 
chiral rotations in $G=SU(3)_L \times SU(3)_R$,
charge conjugation $C$ and parity $P$ are summarized in table
\ref{tab:transformation}.
%
%======================================
%: Table 1
%======================================
%
\begin{table}[t]
\begin{center}
\begin{tabular}{|c|c|c|c|}
\hline \hline
element &
$G$     &
$C$     &
$P$     \\
\hline
$\Sigma$  &
$L \Sigma R^{\dag}$ &
$\Sigma^{T}$ &
$\Sigma^{\dag}$ \\
$\xi$     &
$L \xi U^{\dag} = U \xi R^{\dag}$ &
$\xi^{T}$ &
$\xi^{\dag}$ \\
${\cal O}_{\mu}$     &
$U{\cal O}_{\mu}U^{\dag}$ &
$-O_{\mu}^{T}$ &
${\cal O}_{\mu}$ \\
$S_{\mu}$     &
$S_{\mu}$ &
$-S_{\mu}$ &
$S_{\mu}$ \\
$M$     &
$L M R^{\dag}$ &
$M^{T}$ &
$M^{\dag}$ \\
$M_{\pm}$     &
$U M_{\pm}U^{\dag}$ &
$M_{\pm}^{T}$ &
$\pm M_{\pm}$ \\
$A$     &
$L A R^{\dag}$ &
$A^{T}$ &
$A^{\dag}$ \\
$W_{\pm}$     &
$U W_{\pm}U^{\dag}$ &
$W_{\pm}^{T}$ &
$\pm W_{\pm}$ \\
$V_{\mu}$     &
$UV_{\mu}U^{\dag} + U \partial_{\mu} U^{\dag}$ &
$-V_{\mu}^{T}$ &
$V_{\mu}$ \\
$A_{\mu}$ &
$U A_{\mu}U^{\dag}$ &
$A_{\mu}^{T}$ &
$-A_{\mu}$ \\
${\cal D}_{\nu} {\cal O}_{\mu}$     &
$U({\cal D}_{\nu} {\cal O}_{\mu})U^{\dag}$ &
$-({\cal D}_{\nu} {\cal O}_{\mu})^{T}$ &
${\cal D}_{\nu} {\cal O}_{\mu}$ \\
\hline \hline
\end{tabular}
\caption{
Transformation properties under the group $G$
$=SU(3)_L \times SU(3)_R$, ($L \in SU(3)_L$ and $R \in SU(3)_R$),
charge conjugation $C$ and parity $P$.
In the parity transformed expressions it is
understood that the argument is $(- \stackrel{\rightarrow}{x} ,t)$.
}
\label{tab:transformation}
\end{center}
\end{table}
%======================================

%
%======================================
%3.2
\subsection{Power counting}
\label{subsec:counting}
%======================================
%
Having defined the meson and spurion fields it is straightforward to write down the most general chiral Lagrangian which is compatible with chiral symmetry, charge conjugation and parity.
However, since we have two sources for explicit chiral symmetry breaking, the quark masses and the lattice spacing, their relative size matters for an appropriate power counting. Moreover, the power counting one wants to employ for the vector meson chiral Lagrangian should be consistent with the one adopted for the pure pseudo scalar chiral Lagrangian.

In Ref.\ \cite{Aoki:20051} the
O($a^2$) terms are included in the leading order chiral Lagrangian.
To be consistent we adopt here the following power counting scheme:
\ba
  \mbox{LO}
  &:&
  O(p), O(a),
\non \\
  \mbox{NLO}
  &:&
  O(m), O(a^2), O(ap), O(p^2),
 \label{ListPowerCounting}\\
  \mbox{NNLO}
  &:&
  O(mp), O(p^3), O(am), O(a^3), O(ap^2), O(a^2p).\non
\ea
Here $p$ represents both the residual momentum of the vector meson
and the momentum of the pseudo scalar meson,
and $m$ denotes the quark mass.
This counting scheme assumes $p\sim a$. This implies  $p^2\sim a^2$, and therefore maintains consistency with the power counting of Ref.\ \cite{Aoki:20051} for the pure pseudo scalar sector.

We have listed the NNLO contributions in eq.\ \pref{ListPowerCounting}, because some NNLO terms enter already the one-loop calculation of the vector meson masses. 
To discuss this point, let us define a generic power counting scale $E$
of order $p \sim a$, so that the LO terms or of  O($E$),
the NLO terms are O($E^2$) and so on.
Just from a dimensional analysis one finds that 
a one-loop integral involving a vertex from the LO Lagrangian
gives an O($E^3$) contribution to the vector meson mass.
Therefore, some of the coefficients in the NNLO Lagrangian 
are needed as counterterms for the cancellation of the one-loop divergences.
The coefficients of the NLO Lagrangian, on the other hand,
are not needed as counterterms, since there are no one loop corrections with dimension $E^2$.
This situation is different from the pure pseudo scalar sector
where the divergences in the one-loop contribution from the LO Lagrangian
are canceled by the tree level terms coming from the NLO Lagrangian.

The structure of the correction to the vector meson masses can be summarized as follows:
\ba
  m_V 
  &=&
  \mu_c 
%m_{V_{0}}
  + [\underbrace{
     \mbox{tree level contribution from the LO Lagrangian}}_{{\rm O}(E)}]
\non \\
  &&
\phantom{ \mu_c }  + [\underbrace{
     \mbox{tree level contribution from  the NLO Lagrangian}}_{{\rm O}(E^2)}]
\non \\
  &&
\phantom{ \mu_c }   + [\underbrace{\mbox{one-loop contribution from the LO Lagrangian}}_{{\rm O}(E^3)}]
\non \\
  &&
\phantom{ \mu_c }   + [\underbrace{\mbox{tree level contribution from the NNLO Lagrangian}}_{{\rm O}(E^3)}] + O(E^4),
\label{eqn:massorder}
\ea
where 
%$m_{V_{0}}$ 
$\mu_{c}$ is the vector meson mass in the chiral limit.
In fact,  the second term on the right hand side of eq.\ (\ref{eqn:massorder})
comes from the O($a$) Lagrangian only. As has been shown in Ref.\
\cite{Jenkins:1995vb}, the O($p$) term
does not contribute at tree level
and the chiral correction to the vector meson mass
starts at O($E^2$).

In this paper we are only interested in deriving the vector meson masses through $O(E^3)$. In that case we neither need the NLO terms of  O$(ap,p^2)$ nor the NNLO terms of O$(mp,p^3,ap^2,a^2p)$.
The reason is that all these terms 
contain at least one momentum factor $p$.
By using the lowest order equation of motion,
we can replace the residual vector meson momentum with the momentum $p_{\pi}$ of the pseudo scalar and a contribution to  proportional to $a$.\footnote{The leading order equation of motion, $(-i v \cdot \partial + a \alpha_{\rm s}) S^{\mu}$,  is easily read off from the LO chiral Lagrangian, see eq.\ \pref{EOM} in appendix \ref{app:expLag}. }
The resulting term has a part with at least one $p_{\pi}$
and a part of O$(a^{2},am,a^3)$.
For the former, $p_{\pi} \sim \pal_{\mu} \pi$ implies that this term always contains at least one pseudo scalar field together with a derivative $\pal_{\mu}$.
This kind of term results in a three-point vertex (vector-vector-pseudo scalar)
as a non-vanishing leading term and does not give a tree level contribution to the vector meson mass.
Hence the terms of O$(ap,p^2,mp,p^3,ap^2,a^2p)$ are not needed for our calculation and we do not  list these terms in the next subsection.

%
%=======================================
%3.3
\subsection{Effective Lagrangian}
\label{subsec:lagrangian}
%=======================================
%
In this section, we show our results for the terms in the chiral Lagrangian 
which contribute to the vector meson mass through order $E^3$.
By construction it 
involves the meson fields and 
is invariant under Lorentz
and chiral transformations, charge conjugation and  parity.\footnote{Just for convenience we present our results in Minkowski space, which does not make a difference for the computation of the vector meson masses. The underlying lattice theory, however,  is usually formulated for Euclidean space time.} 

%%%% O($p$)
O($p$):
\vspace{-3mm}
\be
\bay{rcl}
 {\cal L}_{p}
 &=&
 -i S_{\mu}^{\dag} (v \cdot \partial) S^{\mu}\\
 & &
 -i \langle{\cal O}_{\mu}^{\dag} (v \cdot {\cal D}) {\cal O}^{\mu}\rangle
 \\
 & &
 +i g_1
  \Big(S_{\mu}^{\dag} \langle{\cal O}_{\nu} A_{\lambda}\rangle
  -S_{\mu}        \langle{\cal O}_{\nu}^{\dag} A_{\lambda}\rangle\Big)
  v_{\sigma} \epsilon^{\mu \nu \lambda \sigma}\\
  & &
 +i g_2
  \langle\{ {\cal O}_{\mu}^{\dag},{\cal O}_{\nu} \} A_{\lambda}\rangle
  v_{\sigma} \epsilon^{\mu \nu \lambda \sigma},
\eay\label{eqn:p}
\ee
with $\langle X \rangle = \tr ( X )$ for flavor indices.
The derivative $i\pal_{\mu}$ acting on the vector mesons
provides the residual momentum
$r_{\mu}=k_{\mu} - \mu_{V} v_{\mu}$.
We therefore consider the kinetic terms for the vector mesons as O$(p)$.
The terms proportional to  $g_1$ and $g_2$ involve $A_{\lambda}$,
defined in eq.\ \pref{DefAmu}.
Expanding in terms of the pseudo scalar fields this quantity starts as
$A_{\lambda}=\partial_{\lambda}\Pi/f + $O$(\Pi^{3})$. The presence of
the derivative means that both the $g_{1}$ and $g_{2}$ term are of O$(p)$. 

As is well known, the vector mesons are not stable. This effect can be
taken into account by including
an anti-hermitian term in the Lagrangian whose coefficient is
proportional to the decay width.
However, the decay
width is rather small and this contribution is usually ignored \cite{Jenkins:1995vb}.
We also neglect this term in this paper.

%%%% O($a$)
O($a$):
\vspace{-3mm}
\be
\bay{rcl}
  {\cal L}_{a}
  &=&
  \phantom{+}\alpha_1
  \langle 
    W_{+} 
  \rangle 
  S_{\mu}^{\dag} S^{\mu}\\
  &&
 +
  \alpha_2
  \Big(
  \langle
    {\cal O}_{\mu}^{\dag} W_{+}
  \rangle 
  S^{\mu}
 +\langle
    {\cal O}_{\mu} W_{+}
  \rangle 
  S^{\mu \dag}
  \Big)
\\
  &&
 +\alpha_3
  \langle
    \{{\cal O}_{\mu}^{\dag},{\cal O}^{\mu}\} W_{+} 
  \rangle\\
  &&
 +
  \alpha_4
  \langle 
    W_{+}
  \rangle 
  \langle 
    {\cal O}_{\mu}^{\dag}{\cal O}^{\mu}
  \rangle.
\eay\label{eqn:a}
\ee
This O($a$) Lagrangian is a new element of this study. Since the spurion field $A$  associated with the lattice spacing transforms exactly like the quark mass spurion field (cf.\ eq.\  \pref{eqn:spurionA}), this part of the Lagrangian has the same structure as the following O($m$) Lagrangian.

%%%% O($m$)
O($m$):
\vspace{-3mm}
\be
\bay{rcl}
  {\cal L}_{m}
  &=&\phantom{+}
  \lambda_{\rm s}
  \langle 
    M_{+} 
  \rangle 
  S_{\mu}^{\dag} S^{\mu}\\
  & &
 +\lambda_{\rm os}
  (
  \langle
    {\cal O}_{\mu}^{\dag} M_{+}
  \rangle 
  S^{\mu}
 +\langle
    {\cal O}_{\mu} M_{+}
  \rangle 
  S^{\mu \dag}
  )
 \\
  &&
 +\lambda_{\rm o1}
  \langle 
    M_{+}
  \rangle 
  \langle 
    {\cal O}_{\mu}^{\dag}{\cal O}^{\mu}
  \rangle\\
  & & 
 +\lambda_{\rm o2}
  \langle
    \{{\cal O}_{\mu}^{\dag},{\cal O}^{\mu}\} M_{+} 
  \rangle\\
  & & 
  + \Delta\mu\, S_{\mu}^{\dag} S^{\mu}.
\eay\label{eqn:m}
\ee
Note that the leading order mass terms $\mu_{\rm s} S_{\mu}^{\dag} S^{\mu}$
and $\mu_{\rm o} \langle {\cal O}_{\mu}^{\dag} {\cal O}^{\mu} \rangle$,
where $\mu_{\rm s}$ and $\mu_{\rm o}$
are the masses  for the singlet and octet vector meson
in the chiral limit, respectively,
are absent in the $O(p)$ Lagrangian. Their effect is already
included through the phase factor 
$\exp( -im_{V}v\cdot x)$ in the definition of the heavy meson fields.
The mass difference
\bea
\Delta \mu & \equiv & \mu_{\rm s} - \mu_{\rm o}
\ea
between the singlet and octet vector meson mass, however, needs to be
introduced explicitly. This term, which is the last term in eq.\
\pref{eqn:m},  is considered to be of  O$(m)$, since $\Delta \mu < 200$
MeV $\sim m_s$ \cite{Jenkins:1995vb}.\footnote{We assumed the definition
$m_{V}=\mu_{\rm o}$. If we had chosen $m_{V}=\mu_{\rm s}$
the last term in eq.\ \pref{eqn:m} would be
$-\Delta\mu \langle {\cal O}_{\mu}^{\dag} {\cal O}^{\mu} \rangle$. 
Our final results for the vector meson masses are,
of course, independent of this particular choice.} 

%
%===============================================
%%%% O($a^2$)
%===============================================
%
O$(a^2):$
\vspace{-3mm}
\be
\bay{rcl}
  {\cal L}_{a^2}
  &=&\phantom{+ }
  \beta_1
  \langle
    W_{+} W_{+}
  \rangle
  S_{\mu}^{\dag} S^{\mu}\\
  & &
 +\,\beta_2
  \langle 
    W_{+}
  \rangle^{2}
  S_{\mu}^{\dag} S^{\mu}\\
  & &
 +\,\beta_3
  (
  \langle
    {\cal O}_{\mu}^{\dag} W_{+} W_{+}
  \rangle 
  S^{\mu}
 +\langle
    {\cal O}_{\mu} W_{+} W_{+}
  \rangle 
  S^{\mu \dag}
  )
 \\
  &&
 +\,\beta_4
  \langle
    W_{+}
  \rangle
  ( \langle
      {\cal O}_{\mu}^{\dag} W_{+}
    \rangle
     S^{\mu}
  + \langle
      {\cal O}_{\mu} W_{+} 
    \rangle
    S^{\mu\dag})\\
    & &
 +\,\beta_5
  \langle
    \{{\cal O}_{\mu}^{\dag},{\cal O}^{\mu}\} W_{+} W_{+}
  \rangle
 \\
  &&
 +\,\beta_6
  \langle 
    {\cal O}_{\mu}^{\dag} W_{+} {\cal O}^{\mu} W_{+}
  \rangle\\
  & &
 +\,\beta_7
  \langle
    W_{+}
  \rangle
  \langle
    \{ {\cal O}_{\mu}^{\dag}, {\cal O}^{\mu} \} W_{+}
  \rangle\\
  & &
 +\,\beta_8
  \langle 
    W_{+} W_{+}
  \rangle 
  \langle 
    {\cal O}_{\mu}^{\dag}{\cal O}^{\mu}
  \rangle
 \\
  &&
 +\,\beta_9
  \langle
    W_{+}
  \rangle^{2}
  \langle
    {\cal O}_{\mu}^{\dag} {\cal O}^{\mu}
  \rangle\\
  & &
 +\,\beta_{10}
  \langle 
    {\cal O}_{\mu}^{\dag} W_{+} 
  \rangle 
  \langle 
    {\cal O}^{\mu} W_{+}
  \rangle
\\
  &&
 +\,(W_{+} \rightarrow W_{-}).
\eay\label{eqn:asquared}
\ee
%===============================================
In total we find twenty terms, however, the parts involving 
$W_{-}$ are not needed. 
By expanding in powers of the pseudo scalar field one finds $W_{-} = (2i
a/f) \, \Pi$ at leading order. Hence, all these terms have at least two
pseudo scalar fields and therefore do not contribute to the vector meson
masses at tree level.

%===============================================
%%%% O($am$)
%===============================================
O($am$):
\be
\bay{rcl}
  {\cal L}_{am}
  &=&\phantom{+ }
  \gamma_{1}
  \langle
    W_{+} M_{+}
  \rangle
  S_{\mu}^{\dag} S^{\mu}\\
  & &
 +\,\gamma_{2}
  \langle
    W_{+}
  \rangle
  \langle
    M_{+}
  \rangle
  S_{\mu}^{\dag} S^{\mu}\\
  & &
 +\,\gamma_{3}
  \langle
    M_{+}
  \rangle
  (
  \langle
    {\cal O}_{\mu}^{\dag} W_{+}
  \rangle
  S^{\mu}
  +
  \langle
    {\cal O}_{\mu} W_{+}
  \rangle
  S^{\mu\dag}
  )
\\
  &&
 +\,\gamma_{4}
  (
  S_{\mu}^{\dag}
  \langle
    {\cal O}^{\mu} \{ M_{+}, W_{+} \}
  \rangle
  +
  S_{\mu}
  \langle
    {\cal O}^{\mu\dag} \{ M_{+}, W_{+} \}
  \rangle
  )\\
  & &
 +\,\gamma_{5}
  \langle
    W_{+}
  \rangle
  (
  \langle
    {\cal O}_{\mu}^{\dag} M_{+}
  \rangle
  S^{\mu}
  +
  \langle
    {\cal O}_{\mu} M_{+}
  \rangle
  S^{\mu\dag}
  )
 \\
  &&
 +\,\gamma_{6}
  \langle
    M_{+}
  \rangle
  \langle
   \{ {\cal O}_{\mu}^{\dag}, {\cal O}^{\mu} \} W_{+}
  \rangle\\
  & &
 +\,\gamma_{7}
  \langle
    {\cal O}_{\mu}^{\dag} {\cal O}^{\mu}
  \rangle
  \langle
    M_{+} W_{+}
  \rangle\\
  & &
 +\,\gamma_{8}
  \langle
    W_{+}
  \rangle
  \langle
    M_{+}
  \rangle
  \langle
   {\cal O}_{\mu}^{\dag} {\cal O}^{\mu}
  \rangle
\\
  &&
 +\,\gamma_{9}
  \langle
    \{ {\cal O}_{\mu}^{\dag}, {\cal O}^{\mu} \}
    \{ M_{+}, W_{+} \}
  \rangle\\
  & &
 +\,\gamma_{10}
  (
  \langle
    {\cal O}_{\mu}^{\dag}
    M_{+}
    {\cal O}^{\mu}
    W_{+}
  \rangle
  +
  \langle
    {\cal O}_{\mu}^{\dag}
    W_{+}
    {\cal O}^{\mu}
    M_{+}
  \rangle
  )
\\
  &&
 +\,\gamma_{11}
  \langle
    W_{+}
  \rangle
  \langle
   \{ {\cal O}_{\mu}^{\dag}, {\cal O}^{\mu} \} M_{+}
  \rangle\\
  & &
 +\,\gamma_{12}
  (
  \langle
    {\cal O}_{\mu}^{\dag}
    M_{+}
  \rangle
  \langle
    {\cal O}^{\mu}
    W_{+}
  \rangle
  +
  \langle
    {\cal O}_{\mu}^{\dag}
    W_{+}
  \rangle
  \langle
    {\cal O}^{\mu}
    M_{+}
  \rangle
  )
 \\
  &&
 +\,(M_{+}, W_{+} \rightarrow M_{-}, W_{-}).
\label{eqn:am}
\eay
\ee
%===============================================
One can easily check that $M_{-}\sim \Pi$ at leading order.
Therefore, the terms involving $M_{-}$ and $W_{-}$ do not contribute
to the vector meson mass at the order we are working.

For our purposes it is not necessary to list the full
O$(a^3)$ Lagrangian, which is quite cumbersome.
Since we will not encounter any divergences proportional to
$a^3$ in our calculation,
we do not need the coefficients in the O$(a^3)$ Lagrangian as counterterms.
The tree level contribution of the O$(a^3)$ Lagrangian merely gives an
analytic $a^3$ term to the singlet and octet vector meson mass, which we
can simply add to our final one loop result. Note, however, that  the
O$(a^3)$ Lagrangian does not give an off diagonal contribution to the
singlet-octet two point function at O($E^{3})$.
All possible ways to take the trace in terms like
\be
  S_{\mu}^{\dag} {\cal O}^{\mu} W_{+}W_{+}W_{+} + h.c.
\mbox{\hspace{5mm} or\hspace{5mm}}
  S_{\mu}^{\dag} {\cal O}^{\mu} W_{+}W_{-}W_{-} + h.c.
\ee
yield at least two pseudo scalar fields
in the non-vanishing leading term
when one expands $W_{\pm}$ in powers of the $\Pi$ field.
By noting $\langle {\cal O}^{\mu}\rangle=0$,
$W_{+}=1+O(\pi)$ and $W_{-}=O(\pi)$,
one can easily check this statement.

Having determined 
the chiral effective Lagrangian 
we can set the spurion fields to their constant values, i.e.
we take $A \rightarrow a I$ in
eqn.\ \pref{eqn:p} -- \pref{eqn:m}
and eqn.\ \pref{eqn:asquared} -- \pref{eqn:am}. 
This replacement simplifies the fields $W_{\pm}$ to
\bea
W_{\pm}& \rightarrow & \frac{a}{2}(\Sigma\pm \Sigma^{\dag}).
\eea
The Lagrangian in terms of the component fields $\pi^a$, $\rho_{\mu}^a$
for the pseudo scalar and vector mesons is presented in appendix
\ref{sec:feynman}, together with the Feynman rules relevant for the
calculation of the vector meson masses.
%
%=====================================
%3.4
\subsection{Vector meson masses}
\label{subsec:massformula}
%=====================================
%
The full propagator for the $\rho$ and the $K^{\ast}$ is given by
\ba
  i \Delta_{\mu \nu}^{ab}(k)
  &=&
  \frac{-i P_{\mu \nu}}{v \cdot r - a \alpha_{\rm o1}} \delta^{ab}
 +\frac{-i P_{\mu \alpha}}{v \cdot r - a \alpha_{\rm o1}}
  [-i P^{\alpha \beta} \Sigma_{\rm O}^{ab}(k)]
  \frac{-i P_{\beta \nu}}{v \cdot r - a \alpha_{\rm o1}}
 +\cdots
\non \\
  &=&
  -i P_{\mu \nu}
  [v \cdot k - \mu_{\rm o}  - a \alpha_{\rm o1} + \Sigma_{\rm O}(k)]^{-1}_{ab},
\ea
where $a,b$ runs from $1$ to $7$. Recall that 
the external four-momentum for the octet vector meson
is written as $k_{\nu}=\mu_{\rm o} v_{\nu} + r_{\nu}$ 
where $\mu_{\rm o}$ is the octet mass in the chiral limit.
Using the Feynman rules summarized in appendix \ref{sec:feynman}
we obtain for the octet-to-octet self energy through $O(E^3)$
\ba
  \Sigma_{\rm O}^{ab}(k)
  &=&
 -[(N \tilde{M}_0 \lambda_{\rm o1} 
  + a^2 \beta_{\rm o} 
  + N a \tilde{M}_0 \gamma_{\rm o1}) \delta^{ab}
  +(\lambda_{\rm o2} + a \gamma_{\rm o2})
   (2 \tilde{M}_0 \delta_{aa} + \tilde{M}_8 d^{ab8})] 
\nonumber \\
  &&
 +\frac{g_1^2}{f^2}
  \delta^{ab}
  2 C_{21}(w_{\rm s},\tilde{m}_a)
 +\frac{2 g_2^2}{f^2}
  \sum_{c,e}
  d^{ace} d^{bce}
  2 C_{21}(w_{\rm o},\tilde{m}_c)
\non \\
   && 
  +a \frac{2 \alpha_{\rm o1}}{N f^2} \delta^{ab}
   \sum_c I_{0}(\tilde{m}_c)
  +a \frac{2 \alpha_{\rm o2}}{N f^2}
   \sum_{c,e} d^{abe} d^{cce} I_{0}(\tilde{m}_c),\label{Oktet}
\label{eqn:selfenergyforoctet}
\ea
where 
\bea\label{DefScalarVar}
w_{\rm o,s}& = & 
v \cdot r - a \alpha_{\rm o1,s},
\eea 
and $d^{abc}$ is the totally symmetric $d$--symbol of
su($N$).\footnote{We keep the number of flavors $N$ undetermined at
intermediate stages since $N$ provides some useful checks for our
equations. In the final results we will set $N$ equal  to 3. Note,
however, that intermediate results like eq.\ \pref{Oktet} do not hold
for arbitrary $N$ unless all masses are degenerate.}
The coefficients $\alpha_{{\rm o}i}$ are particular combinations of the
low energy constants $\alpha_{1}, \ldots, \alpha_{4}$ which enter eq.\
\pref{eqn:a}. Similarly, the coefficients $\beta_{\rm o}$ and
$\gamma_{{\rm o}i}$
are combinations of the low energy constants
$\beta_{1},\ldots,\beta_{10}$ and $\gamma_{1},\ldots,\gamma_{12}$,
respectively. The full expressions for these combinations can be found
in appendix \ref{sec:feynman}, even though it will not be necessary to keep
track of the original constants in the following.
The functions $C_{21}(\omega,M)$ and $I_{0}(M)$ stem from the loop
integral and are defined as
\ba
  C_{21}(w,M)
  &=&
  \frac{1}{3}
  \left[
    (M^2 - w^2) J(w,M) + w I_0(M)
  \right]
 -\frac{w}{12 \pi^2}
  \left( 
    \frac{M^2}{2} - \frac{w^2}{3}
  \right),\label{FunctionC21}
\\
  J(w,M)
  &=&
  \frac{w}{8 \pi^2}
  \left[
    R + \ln \frac{M^2}{\mu^2} -1
  \right]\nonumber\\
& &  + \frac{1}{8 \pi^2}
  \left[
    2 \sqrt{M^2-w^2} \mbox{arccos}
    \left(
      -\frac{w}{M}
    \right)
  \right],\quad
\mbox{\hspace{2mm}for\hspace{2mm}}
  w^2 < M^2,
\label{eqn:J}
\\[2ex]
  I_{0}(M)
  &=&
  \frac{M^2}{16 \pi^2}
  \left[
    R + \ln \frac{M^2}{\mu^2}
  \right],
\\
  R
  &=&
  \frac{2}{n-4}
 -[ \ln(4 \pi) + \Gamma^{\prime}(1) + 1],\label{FunctionR}
\ea
with the dimension $n$.
The one-loop integrals encountered here are essentially the same as those in 
baryon chiral perturbation theory, and a 
detailed derivation of the formulae \pref{FunctionC21} -  \pref{FunctionR}
can be found in Ref.\ \cite{Scherer:2002tk}.

In order to cancel the divergence in $R$, defined in eq.\ \pref{FunctionR},
the O($am$) coefficients $\gamma_{i}$ $(i={\rm o1, o2})$
need to be properly renormalized,
\be\label{RenCond}
  \gamma_i
  =
  \gamma_i^r(\mu) + \frac{2 B C_i}{8 \pi^2 f^2} R,
\ee
where 
the coefficient $B$ is a low-energy parameter in the LO chiral Lagrangian
for the pseudo scalars  (see Ref.\ \cite{Aoki:20051} and appendix 
\ref{sec:feynman}).
The coefficients $\gamma_i^r(\mu)$ denote renormalized parameters,
$\mu$ is the renormalization scale\footnote{In the following we will
simply write $\gamma_i^r$ for the renormalized parameters and suppress
the dependence on $\mu$.},
and the coefficients $C_i$ are given by
\ba
  C_{\rm o1} 
  &=&
  \alpha_{\rm o1}
  \frac{N^2-1}{ N}  - \alpha_{\rm o2}
  \frac{5}{3 N},
\label{eqn:finiteparto1}
\\
  C_{\rm o2}
  &=&
  \alpha_{\rm o2}
  \frac{5}{6 N}.
\label{eqn:finiteparto2}
\ea

The on-shell condition for the $\rho$ and $K^{\ast}$ meson
($a=1,\cdots,7$) is given by (no sum over $a$)
\ba
\begin{array}{rcl}
  0
  &=&
  v \cdot k - \mu_{\rm o} - a \alpha_{\rm o1} + \Sigma_{\rm O}^{aa}(k),
\\
  k_{\mu}
  &=&
  m_{a} v_{\mu},
\qquad
  k^2 = m_{a}^2 ,
\label{eqn:onshellcondition}
\end{array}
\ea
where $m_{a}$ is the physical mass. 
So far we have chosen to express the self energy
as a function of the four-momentum $k_{\mu}$.
$\Sigma_{\rm O}^{aa}(k)$ may also be parametrized by the scalar variables
$w_{\rm o}$ and $w_{\rm s}$, defined in eq.\ \pref{DefScalarVar}, and we can
write $\Sigma_{\rm O}^{aa}(k) = \Sigma_{\rm O}^{aa}(w_{\rm o},w_{\rm s})$.
%Their dependence comes from $C_{21}$.
It is easy to see that the definition of $w_{\rm o}$ together with the on-shell condition \pref{eqn:onshellcondition} yields,
\ba
w_{\rm o}
&=&
m_a - \mu_{\rm o} - a\alpha_{\rm o1} \,=\,
- \Sigma^{aa}_{\rm O}(k=m_a v).
\ea
%The last equal is a consequence of the
%on-shell condition in eq.(\ref{eqn:onshellcondition}).
$w_{\rm o}$ is considered to be of  $O(E^2)$
since the self energy $\Sigma^{aa}_{\rm O}$ is of $O(E^2)$.
For $w_{\rm s}$ we find the similar result 
\ba
w_{\rm s}
&=&
m_a - \mu_{\rm o} - a\alpha_{\rm s} \,=\,
- \Sigma^{aa}_{\rm O}(k=m_a v) - a(\alpha_{\rm s}-\alpha_{\rm o1}).
\ea
%and the leading order is $O(a)$ but not $O(E^2)$.
Here we assume  that 
the difference between the leading lattice artifacts for the 
octet and the singlet mass
is of order $E^2$ and therefore small\footnote{
If this assumption is not valid,
then $w_{\rm s}$ is of $O(a)\sim O(E)$.
In this case, the function $C_{21}(w_s,\tilde m_a)$ in eq.(\ref{eqn:selfenergyforoctet})
contains divergences whose coefficients are of order $a m$ and $a^3$.
Such divergences would alter the renormalization conditions for $\gamma_{\rm o1, o2}$ in eq. \pref{RenCond}.
%which is the low energy constants of the $O(am)$ terms,
%may be different from
%eq.(\ref{eqn:finiteparto1}, \ref{eqn:finiteparto2})
%where it is assumed that $a(\alpha_{\rm s}-\alpha_{\rm o1})\sim O(E^2)$ holds
Also the $O(a^3)$ counterterms may receive a divergent renormalization in order to cancel additional divergences.
Furthermore, if $w_{\rm s}$ is of $O(a)$,
the condition $w^2<M^2$ in eq.\ (\ref{eqn:J}) may not be valid.
Hence the functional form of $J(w,M)$ might be changed
according to the relation between $w$ and $M$ as
in Ref.\ \cite{Scherer:2002tk} eq.(C.27).
},
\be\label{AssumptionLattArtifacts}
  a (\alpha_{\rm s}
-    \alpha_{\rm o1})
  =
  a \Delta \alpha
  =
  O(E^2).
\ee
This is supported by a strong coupling analysis performed in Ref.\ \cite{Aoki:1986kt}.
%where it was shown that in the second order of effective potential 
Expanding the effective potential through second order in terms of the vector meson fields, 
it has been shown that there exists no
difference between the singlet and octet vector meson to all orders in the strong coupling expansion.
Whether  $w_{\rm s}$ and $w_{\rm o}$ is small away from the strong coupling limit can and should be checked in the actual lattice simulation.

Provided the assumption \pref{AssumptionLattArtifacts}, both $w_{\rm s}$ and $w_{\rm o}$  are of
O($E^2$) and we can set 
$w_{\rm o,s} = 0$ in $\Sigma_{\rm O}^{aa}$, since the difference is
beyond the order we are working to. We therefore conclude that 
\be
  m_a 
  = 
  \mu_{\rm o} + a \alpha_{\rm o1}
- \Sigma_{\rm O}^{aa}(w_{\rm o}=0,w_{\rm s}=0).
\ee
Note that the function 
$C_{21}$ becomes much simpler when the first argument is set to zero, 
\bea
C_{21}(w=0,M) & = &  \frac{1}{24 \pi} M^3.
\eea

Let us now consider the sector of the $\rho^{8}$ and the $\phi^{0}$ mesons.
Due to mixing,
the propagator for this sector is written as a 2$\times$2 matrix,
\be
  i \Delta_{\mu \nu}(k)
  =
  -i P_{\mu \nu}
  \left[
  \begin{array}{cc}
  v \cdot r  - a \alpha_{\rm o1} + \Sigma_{\rm O}(k)   & \Sigma_{\rm OS}(k)\\
  \Sigma_{\rm OS}(k) &   v \cdot r  - a \alpha_{\rm s} + \Sigma_{\rm S}(k)
  \end{array}
  \right]^{-1},
\ee
where the self energies for octet-to-octet ($a=8$), singlet-to-singlet
and octet-to-singlet are given by
\ba
  \Sigma_{\rm O}(k)
  &=&
  \Sigma_{\rm O}^{a=8,b=8}(k), \\
  \Sigma_{\rm S}(k)
  &=&
 -(N \tilde{M}_0 \lambda_{\rm s}
  + a^2 \beta_{\rm s}
  + N a \tilde{M}_0 \gamma_{\rm s}
  + \Delta\mu)
\nonumber \\
  && 
 +\frac{g_1^2}{f^2}
  \sum_{c}
  2 C_{21}(w_{\rm o},\tilde{m}_c)
 +a \frac{2 \alpha_{\rm s}}{N f^2}
  \sum_c I_0(\tilde{m}_c),
\\
  \Sigma_{\rm OS}(k)
  &=&
 -\frac{\tilde{M}_8}{\sqrt{2}} (\lambda_{\rm os} + a \gamma_{\rm os} )
\nonumber \\
  &&
 +\frac{\sqrt{2} g_1 g_2}{f^2}
  \sum_{c}
  d^{cc8}
  2 C_{21}(w_{\rm o},\tilde{m}_c)
 + a \frac{\alpha_{\rm os} \sqrt{2}}{Nf^2}
  \sum_{c}
  d^{cc8}
  I_0 (\tilde{m}_c).
\ea
For the cancellation of $R$,
in a similar way to the case of $\rho$-$K^{\ast}$ sector,
O($am$) coefficients $\gamma_{i}$ $(i={\rm s, os})$
should be renormalized as in eq.(\ref{RenCond}) with
\ba
  C_{\rm s}
  &=&
  \alpha_{\rm s}
  \frac{N^2-1}{N^2},
\\
  C_{\rm os}
  & =&
  \alpha_{\rm os}
  \frac{5}{6 N}.
\ea
The physical masses $m_{i},\,i=1,2,$ for this sector are determined by the on-shell condition
\be
 {\rm det}
 \left[
  \begin{array}{cc}
   v \cdot r - a \alpha_{\rm o1} + \Sigma_{\rm O}(k) & \Sigma_{\rm OS}(k) \\
   \Sigma_{\rm OS}(k) & v \cdot r - a \alpha_{\rm s}+ \Sigma_{\rm S}(k)
  \end{array}
 \right]
 =0,
\ee
at $k_{\mu}=m_{i} v_{\mu}$.
The task is to solve the equation
\be
 {\rm det}
 \left[
  \begin{array}{cc}
   m - \mu_{\rm o} - a \alpha_{\rm o1} + \Sigma_{\rm O}(k=mv) & 
   \Sigma_{\rm OS}(k=mv) \\
   \Sigma_{\rm OS}(k=mv) & 
   m - \mu_{\rm s} - a \alpha_{\rm s}+ \Sigma_{\rm S}^{\prime}(k=mv)
  \end{array}
 \right]
 =0,
\ee
with respect to the mass $m_{i}$,
where we have defined $\Sigma_{\rm S}^{\prime}$ as
$\Sigma_{\rm S}=\Sigma_{\rm S}^{\prime} - \Delta\mu$.
The solutions of the equation are given by
\ba
  m_{\pm}
  &=&
  \frac{1}{2}
  \daileft
    \mu_{\rm s} + \mu_{\rm o}
  + a (\alpha_{\rm s} + \alpha_{\rm o1})
  - (\Sigma_{\rm S}^{\prime} + \Sigma_{\rm O})
\non \\ & &
  \pm
  \sqrt{
    \left\{
    \Delta \mu + a \Delta \alpha
    - (\Sigma_{\rm S}^{\prime} - \Sigma_{\rm O})
    \right\}^2
  + 4 (\Sigma_{\rm OS})^2
  }
  \dairight.
\ea
Assuming that the SU(3) breaking $(\tilde m - \tilde m_s)/3$
is smaller than the average $(2\tilde m + \tilde m_s)/3$ of the quark masses\footnote{This assumption means
that the diagonal part is dominant compared to the off-diagonal
element,
$\{\Delta\mu+a\Delta\alpha-(\Sigma_{\rm S}^{\prime}-\Sigma_{\rm
 O})\}^2>4(\Sigma_{\rm OS})^2$} together with
$\{\Delta\mu+a\Delta\alpha-(\Sigma_{\rm S}^{\prime}-\Sigma_{\rm O})\}>0$,
the physical masses are given by
\ba
  m_{+}
  &=&
  \mu_{\rm s} + a \alpha_{\rm s} - \Sigma_{\rm S}^{\prime}
+ \frac{(\Sigma_{\rm OS})^2}{
    \Delta \mu + a \Delta \alpha
    - (\Sigma_{\rm S}^{\prime} - \Sigma_{\rm O})
  },
\\
  m_{-}
  &=&
  \mu_{\rm o} + a \alpha_{\rm o1} - \Sigma_{\rm O}
- \frac{(\Sigma_{\rm OS})^2}{
    \Delta \mu + a \Delta \alpha
    - (\Sigma_{\rm S}^{\prime} - \Sigma_{\rm O})
  }.
\ea
As in the case of the $\rho$ and $K^{\ast}$ sector,
we consider the self energies $\Sigma_{\rm O,OS}$ and $\Sigma_{\rm S}^{\prime}$ as functions of$w_{\rm o,s}$ when we impose the on-shell condition.
%in the $\Sigma_{\rm O,OS}$ and $\Sigma_{\rm S}^{\prime}$.
Setting the physical masses equal to $m_{\pm}$,
the parameters $w_{\rm o,s}$ are given as
\ba
  w_{\rm o}
  &=&
  m - \mu_{\rm o} - a\alpha_{\rm o1}
\non \\
  &=&
  \left\{
  \begin{array}{ll}
  \Delta \mu + a \Delta \alpha
  - \Sigma_{\rm S}^{\prime}
  - \frac{(\Sigma_{\rm OS})^2}{
    \Delta \mu + a \Delta \alpha
    - (\Sigma_{\rm S}^{\prime} - \Sigma_{\rm O})
  },
  & \mbox{  for  } m=m_{+},
  \\
  - \Sigma_{\rm O}
  - \frac{(\Sigma_{\rm OS})^2}{
      \Delta \mu + a \Delta \alpha
    - (\Sigma_{\rm S}^{\prime} - \Sigma_{\rm O})
  },
  & \mbox{  for  } m=m_{-},
  \end{array}
  \right.
\\
  w_{\rm s}
  &=&
  m - \mu_{\rm o} - a\alpha_{\rm s}
\non \\
  &=&
  \left\{
  \begin{array}{ll}
  \Delta \mu
  - \Sigma_{\rm S}^{\prime}
  - \frac{(\Sigma_{\rm OS})^2}{
    \Delta \mu + a \Delta \alpha
    - (\Sigma_{\rm S}^{\prime} - \Sigma_{\rm O})
  },
  & \mbox{  for  } m=m_{+},
  \\
  - a \Delta \alpha
  - \Sigma_{\rm O}
  - \frac{(\Sigma_{\rm OS})^2}{
      \Delta \mu + a \Delta \alpha
    - (\Sigma_{\rm S}^{\prime} - \Sigma_{\rm O})
  },
  & \mbox{  for  } m=m_{-}.
  \end{array}
  \right.
\ea
Hence it turns out that
both $ w_{\rm o}$ and $ w_{\rm s}$ are of $O(E^2)$,
since $\Delta\mu$, $a\Delta\alpha$, $\Sigma_{\rm O,S}$
and $\Sigma_{\rm OS}$ are of $O(E^2)$.
Therefore, we can set $w_{\rm o,s}=0$ in the self energies,
as we did in the $\rho$--$K^{\ast}$ sector.

With these preparations our one-loop results for the vector meson masses
are as follows:

%======================
%:final result
%======================
\ba
  m_{\rho}
  &=&
  m_{\rm O}(a)
+ \lambda_x(a) x + 2 \lambda_y(a) y
\non \\
  &-&
  \frac{1}{12 \pi f^2}
  \daileft
    (g_1^2 + \frac{2}{3} g_2^2) (x+2y)^{\frac{3}{2}}
  + 2 g_2^2 (x-y)^{\frac{3}{2}}
  + \frac{2}{3} g_2^2 (x-2y)^{\frac{3}{2}}
  \dairight
\non \\
  &-&
  \frac{2a}{3f^2}
  \daileft
    \shouleft 
      3\alpha_{\rm o1} + \alpha_{\rm o2}
    \shouright 
    L_{\pi}
  + \shouleft
      4\alpha_{\rm o1} - \frac{2}{3} \alpha_{\rm o2}
    \shouright 
    L_{K}
  + \shouleft
      \alpha_{\rm o1} - \frac{1}{3} \alpha_{\rm o2}
    \shouright 
    L_{\eta}
  \dairight,
\\
  m_{K^{\ast}}
  &=&
  m_{\rm O}(a)
+ \lambda_x(a) x - \lambda_y(a) y
\non \\
  &-&
  \frac{1}{12 \pi f^2}
  \daileft
    \frac{3}{2} g_2^2 (x+2y)^{\frac{3}{2}}
  + \left( g_1^2 + \frac{5}{3} g_2^2 \right) (x-y)^{\frac{3}{2}}
  + \frac{1}{6} g_2^2 (x-2y)^{\frac{3}{2}}
  \dairight
\non \\
  &-&
  \frac{2a}{3f^2}
  \daileft
    \shouleft 
      3\alpha_{\rm o1} - \frac{1}{2} \alpha_{\rm o2}
    \shouright 
    L_{\pi}
  + \shouleft
      4\alpha_{\rm o1} + \frac{1}{3} \alpha_{\rm o2}
    \shouright 
    L_{K}
  + \shouleft
      \alpha_{\rm o1} + \frac{1}{6} \alpha_{\rm o2}
    \shouright 
    L_{\eta}
  \dairight,
\\
  m_{+}
  &=&
  m_{(00)}
+ \frac{m_{(08)}^2}{m_{(00)}-m_{(88)}},
\\
  m_{-}
  &=&
  m_{(88)}
- \frac{m_{(08)}^2}{m_{(00)}-m_{(88)}},
\\
  m_{(88)}
  &=&
  m_{\rm O}(a)
+ \lambda_x(a) x - 2 \lambda_y(a) y
\non \\
  &-&
  \frac{1}{12 \pi f^2}
  \daileft
    2 g_2^2 (x+2y)^{\frac{3}{2}}
  + \frac{2}{3} g_2^2 (x-y)^{\frac{3}{2}}
  + \left( g_1^2 + \frac{2}{3} g_2^2 \right) (x-2y)^{\frac{3}{2}}
  \dairight
\non \\
  &-&
  \frac{2a}{3f^2}
  \daileft
    \shouleft 
      3\alpha_{\rm o1} - \alpha_{\rm o2}
    \shouright 
    L_{\pi}
  + \shouleft
      4\alpha_{\rm o1} + \frac{2}{3} \alpha_{\rm o2}
    \shouright 
    L_{K}
  + \shouleft
      \alpha_{\rm o1} + \frac{1}{3} \alpha_{\rm o2}
    \shouright 
    L_{\eta}
  \dairight, \\
  m_{(00)}
  &=&
  m_{\rm S}(a)
+ \sigma_x(a) x
\non \\
  &-&
  \frac{1}{12 \pi f^2}
  g_1^2
  \daileft
    3 (x+2y)^{\frac{3}{2}}
  + 4 (x-y)^{\frac{3}{2}}
  + (x-2y)^{\frac{3}{2}}
  \dairight
\non \\
  &-&
  \frac{2a}{3f^2}
  \alpha_{\rm s}
  \daileft 
    3 L_{\pi}
  + 4 L_{K}
  + L_{\eta}
  \dairight,
\\
  m_{(08)}
  &=&
  \sigma_y(a) y
\non \\
  &-&
  \frac{1}{12 \pi f^2}
  g_1 g_2
  \sqrt{\frac{2}{3}}
  \daileft
    3 (x+2y)^{\frac{3}{2}}
  - 2 (x-y)^{\frac{3}{2}}
  - (x-2y)^{\frac{3}{2}}
  \dairight
\non \\
  &-&
  \frac{a}{3f^2}
  \alpha_{\rm os}
  \sqrt{\frac{2}{3}}
  \daileft 
    3 L_{\pi}
  - 2 L_{K}
  - L_{\eta}
  \dairight,\\
  \tan 2 \theta
&  =&
  \frac{2 m_{(08)}}{m_{(00)}-m_{(88)}},
\label{eqn:vectormesonmass}
\ea
%======================
%end final result
%======================
where $\theta$ denotes the mixing angle in the $\rho^{8}$--$\phi^{0}$ sector.
We introduced the two parameters
\ba
  x
  &=&
  \frac{2B}{3}
  (2 \tilde{m} + \tilde{m}_s),\\[2ex]
  y
  &=&
  \frac{B}{3}
  (\tilde{m} - \tilde{m}_s),
\ea
as a short hand notation for convenient combinations of
quark masses \cite{Aoki:20051}. 
The leading order pseudo scalar masses assume a very simple form in
terms of $x$ and $y$,
 \be
 \bay{rcl}
  \tilde{m}_{\pi}^{2}
  &=&
  x + 2y, \\ 
  \tilde{m}_{K}^{2} & =& x - y,\\
\tilde{m}_{\eta}^{2}&  =&
  x - 2y\,.
  \eay
  \ee
For the chiral logarithms stemming from the loop integration we have introduced    
 \ba
  L_{\pi}
  &=&
  \frac{\tilde{m}_{\pi}^2}{16 \pi^2}
  \ln \left(\frac{\tilde{m}_{\pi}^2}{\mu^2}\right),\nonumber\\
  L_{K}
  &=&
  \frac{\tilde{m}_{K}^2}{16 \pi^2}
  \ln \left(\frac{\tilde{m}_{K}^2}{\mu^2}\right),\\
  L_{\eta}
  &=&
  \frac{\tilde{m}_{\eta}^2}{16 \pi^2}
  \ln \left(\frac{\tilde{m}_{\eta}^2}{\mu^2}\right).\nonumber
\ea
The functions
\ba
  m_{\rm O}(a)
  &=&
  V_0 + V_1 a + V_2 a^2 + V_3 a^3,
  \\
  m_{\rm S}(a)
  & = &
  S_0 + S_1 a + S_2 a^2 + S_3 a^3, \\
  \lambda_x(a)
  &=&
  \lambda_{x}^{(0)} + \lambda_{x}^{(1)} a,
\mbox{\hspace{10mm}}
  \lambda_y(a)
  =
  \lambda_{y}^{(0)} + \lambda_{y}^{(1)} a,\\
  \sigma_x(a)
  &=&
  \sigma_{x}^{(0)} + \sigma_{x}^{(1)} a,
\mbox{\hspace{10mm}}
  \sigma_y(a)
  =
  \sigma_{y}^{(0)} + \sigma_{y}^{(1)} a,
 \label{eqn:parameters}
\ea
parametrize the analytic lattice spacing dependence.  
We have, for convenience and for the sake of transparency, introduced
  new constants $V_{i},S_{i},\, i=1,\dots4$ and
  $\lambda_{x,y}^{(j)},\,\sigma_{x,y}^{(j)},\,j=1,2$. These new
  constants are combinations of the low energy constants in the chiral
  Lagrangian. They are explicitly given
\ba
V_0&=&\mu_{\rm o},
\qquad
V_1=\alpha_{\rm o1},
\qquad
V_2=\beta_{\rm o},
\\
S_0&=&\mu_{\rm s},
\qquad
S_1=\alpha_{\rm s},
\qquad
S_2=\beta_{\rm s},
\\
\lambda_{x}^{(0)}&=& 
\frac{N \lambda_{\rm o1}    + 2 \lambda_{\rm o2}}{2B},
\qquad
\lambda_{x}^{(1)}= 
\frac{N \gamma_{\rm o1}^{r} + 2 \gamma_{\rm o2}^{r}}{2B},
\\
\lambda_{y}^{(0)}&=&
\frac{\lambda_{\rm o2}}{B},
\qquad
\lambda_{y}^{(1)}=
\frac{\gamma_{\rm o2}^{r}}{B},
\\
\sigma_{x}^{(0)}&=& 
\frac{N \lambda_{\rm s}}{2B},
\qquad
\sigma_{x}^{(1)}= 
\frac{N \gamma_{\rm s}^{r}}{2B},
\\
\sigma_{y}^{(0)}&=&
\frac{\sqrt{6}\lambda_{\rm os}}{B},
\qquad
\sigma_{y}^{(1)}=
\frac{\sqrt{6}\gamma_{\rm os}^{r}}{B}.
\ea
However, from a
practical point of view there is no need to keep track of the original
low energy constants.  
Note that we
have added an O$(a^3)$ term for the octet and singlet
vector meson mass,  but not for the singlet-octet mixed contribution $m_{(08)}$, 
as discussed in sect.\ \ref{subsec:lagrangian}.

Note that there are
chiral logarithms present 
in the results for the vector meson masses.
These chiral corrections are a lattice artifact, as can be seen from the
presence of the $a$ in front of the logarithms.
This chiral log contribution vanishes in the continuum limit
and the mass formulae converge to the results given
in Ref.\ \cite{Jenkins:1995vb}.
However, as long as the lattice spacing is non-zero, there are two kinds
of non-analytical quark mass dependence, proportional to
$m_{q}^{3/2}$ and $m_{q} \log m_{q}$. It will be very interesting to study
the competition of these two contributions in actual lattice simulations.

Let us count the number of unknown parameters. 
In the case of the $\rho$ and $K^{\ast}$ meson, we find twelve unknown
combinations of low-energy constants:
$V_{i}, \,\lambda_{x}^{(j)},\,\lambda_{y}^{(j)}$, $g_1$, $g_2$,
$\alpha_{\rm o1}$, $\alpha_{\rm o2}$.
However, as long
as we are interested in performing fits to lattice data at a given and
fixed lattice spacing $a$, this number essentially reduces to seven
independent constants:
$m_{\rm O}$, $\lambda_x$, $\lambda_y$,
$g_1$, $g_2$, $\alpha_{\rm o1}$, $\alpha_{\rm o2}$.
For the $m_{\pm}$ sector,
we have additional ten low energy constants:
$S_{i}, \,\sigma_{x}^{(j)},\,\sigma_{y}^{(j)}$,
$\alpha_{\rm os}$, $\alpha_{\rm s}$.
When we consider at a fixed lattice spacing,
there are essentially five independent constants:
$m_{\rm S}$, $\sigma_x$, $\sigma_y$,
$\alpha_{\rm os}$, $\alpha_{\rm s}$.

%4
%=============================================
\section{Concluding remarks} 
\label{sec:Conclusion}
%=============================================
%
We have derived the one loop expressions for vector meson masses using
an effective theory based on the heavy vector meson formalism of Ref.\
\cite{Jenkins:1995vb}. The effects due to a non-vanishing lattice
spacing introduce a fair number of new unknown low-energy constants in
the chiral Lagrangian. However,  the actual number of fit parameters in
the mass formulae seems small enough for the expressions to be useful
for the chiral extrapolation of actual lattice data. In the case of the
$\rho$ and $K^{\ast}$ meson mass, the number of unknown fit parameters
is seven. 
The numerical simulations of the CP-PACS/JLQCD
collaboration are carried out with
five different (degenerate) up and down quark masses and two different
values for the strange quark mass. The total number of independent data
points is 18 without data for the $\phi$ and $\omega$ mesons. If one
measures these masses including noisy disconnected contributions, the number of data points increases to 36. In both cases the number of data points is well above 7.

Here we have calculated the expressions for the vector meson masses
only. In the calculation of the vector meson decay constants some
additional low-energy constants enter, depending on the choice for the
vector current. Apart from this the calculation is straightforward and
currently under way \cite{Aoki:20054}.

Our main motivation for the calculation presented here was to derive fit
forms for the chiral extrapolation of the vector meson masses.
If our formulae describe the lattice QCD data well,
we obtain, as a byproduct,
estimates for the low energy constants of the chiral 
effective theory for vector mesons.
Particularly promising are the leading order parameters $g_1$ and $g_2$. 
These couplings are not corrected by lattice artifacts through O($E^3$).
Provided that the corrections of O($E^4$) and higher are negligible,
we have a good chance to determine $g_1$ and $g_2$ in a fit, 
even if lattice data is available for one lattice spacing only.
The chiral quark model \cite{Manohar:1983md} predicts $g_2=0.75$
in the large $N_c$ limit where
$g_1 =  \frac{2}{\sqrt{3}} g_2$.
On the other hand, the estimate $g_2 \sim 0.6$ is given in
Ref.\ \cite{Davoudiasl:1995ed}, where
vector meson  \chpt\  is applied to $\tau$ decay processes together with
a comparison of the results with experimental data. 
If we are able to extract $g_1$ and $g_2$ in a fit to lattice QCD data, 
we obtain results based on first principles QCD without relying on model
dependent assumptions or the large $N_{c}$ limit.

The heavy vector meson formalism of Ref.\ \cite{Jenkins:1995vb} is not
the only way to derive an effective theory for vector mesons. Another
approach employs the so-called hidden local symmetry \cite{Harada:2003}.
Also this approach has a
definite counting scheme, and 
in principle one can formulate it for non-zero lattice spacing too. It
would be interesting to perform the calculation of the vector meson
masses based on this alternative approach. The comments in the previous
paragraph about the extraction of low energy constants through a fit to
lattice data applies to this effective theory as well. 

%
%
%=====================
\section*{Acknowledgments}
%=====================
%
This work is supported in part by the Grants-in-Aid for
Scientific Research from the Ministry of Education, 
Culture, Sports, Science and Technology 
(Nos. 13135204, 15204015, 15540251, 16028201,
16$\cdot$11968%Takeda
). 
O.\ B.\ is supported in part by the University of Tsukuba Research Project.
S.\ T.\ is supported by Research Fellowships of
the Japan Society for the Promotion of Science for Young Scientists.

%
%=======
\appendix
%=======
%
%=====================
\section{Feynman rules}
\label{sec:feynman}
%=====================
%
In this appendix, we present the Feynman rules needed in the calculation for the vector meson masses.
The lattice spacing $a$ shown in the following equations stems from the spurion field $W_{\pm}$ once the replacement $A\rightarrow a I$ has been made.
%
%==========================
\subsection{Expanding the Lagrangian}
\label{app:expLag}
%==========================
%
In terms of the component fields,
the LO Lagrangian which includes up to two 
pseudo scalar meson fields is expressed as
\ba
  {\cal L}_{\rm LO}
  &=&
  {\cal L}_{p} + {\cal L}_{a}
\non \\
  &=&
  ({\cal L}_{p}^{0\pi} + {\cal L}_{a}^{0\pi})
+ {\cal L}_{p}^{1\pi}
+ ({\cal L}_{p}^{2\pi} + {\cal L}_{a}^{2\pi})
+ O(\pi^3),
\\
  {\cal L}_{p}^{0\pi} + {\cal L}_{a}^{0\pi}
  &=&
  S_{\mu}^{\dag} (-i v \cdot \partial + a \alpha_{\rm s}) S^{\mu}
 +\sum_a 
  \rho_{\mu}^{a \dag} (-i v \cdot \partial + a \alpha_{\rm o1}) \rho^{a \mu},\label{EOM}
\\
  {\cal L}_{p}^{1\pi}
  &=&
  \frac{i g_1}{f} \sum_a
  (S_{\mu}^{\dag} \rho_{\nu}^{a}-S_{\mu} \rho_{\nu}^{a \dag})
  \partial_{\lambda} \pi_a
  v_{\sigma} \epsilon^{\mu \nu \lambda \sigma}
  +
  \frac{i g_2 \sqrt{2}}{f} \sum_{a,b,c}
  d_{abc} \rho_{\mu}^{a \dag} \rho_{\nu}^{b} 
  \partial_{\lambda} \pi_c
  v_{\sigma} \epsilon^{\mu \nu \lambda \sigma},
\\
  {\cal L}_{p}^{2\pi} + {\cal L}_{a}^{2\pi}
  &=&
 -\frac{i}{f^2} \sum_{a,b,c,d,e}
  f_{ade} f_{bce} 
  \rho_{\mu}^{a \dag} \pi^b (v \cdot \partial \pi^c) \rho^{d \mu}
\non \\
  &&
 -a \frac{\sqrt{2}\alpha_{\rm os}}{Nf^2}
  \sum_{a,b,c} d_{abc}
  (S^{\mu}\rho_{\mu}^{a\dag} \pi^b \pi^c
  +S^{\mu\dag}\rho_{\mu}^{a} \pi^b \pi^c)
\nonumber \\
  &&
 -a \frac{2}{Nf^2}
  \sum_{a,b,c,d}
  (\alpha_{\rm o1} \delta_{ab} \delta_{cd}
  +\alpha_{\rm o2} \sum_e d_{abe} d_{cde})
  \rho_{\mu}^{a\dag}\rho^{\mu b} \pi^c \pi^d
\nonumber \\
  &&
 -a \frac{2 \alpha_{\rm s}}{N f^2} 
  \sum_{a,b}
  \delta_{ab}
  S_{\mu}^{\dag} S^{\mu} \pi^a \pi^b,
\ea
where the newly introduced coefficients relate to the ones defined in eq.\ \pref{eqn:a} according to
\ba
  \alpha_{\rm s}
  &=&
  N \alpha_1,\\
%\mbox{\hspace{10mm}}
  \alpha_{\rm os}
&  = &
  N \alpha_2,
\\
  \alpha_{\rm o1}
  &=&
  2 \alpha_3 + N \alpha_4,\\
% \mbox{\hspace{10mm}}
  \alpha_{\rm o2}
  & = &
  N \alpha_3.
\ea
The summation over roman indices is understood to range from 1 to 8.
$f_{abc}$ and $d_{abc}$ are the structure constants and the totally
symmetric $d$--symbol of su(N) respectively. Obviously, we are most
interested in the case $N=3$.

The NLO and NNLO Lagrangian results in the following tree level
contribution to the vector meson masses:
\ba
  {\cal L}_{\rm NLO} + {\cal L}_{\rm NNLO}
  &=&
  ({\cal L}_{m} + {\cal L}_{a^2}) + {\cal L}_{am}
\non \\
  &=&
  {\cal L}_{m}^{0\pi} + {\cal L}_{a^2}^{0\pi} + {\cal L}_{am}^{0\pi}
  + O(\pi),
\\
  {\cal L}_{m}^{0\pi} + {\cal L}_{a^2}^{0\pi} + {\cal L}_{am}^{0\pi}
  &=&
  (N \tilde{M}_0 \lambda_{\rm s}
 + a^2 \beta_{\rm s}
 + N \tilde{M}_0 a \gamma_{\rm s}
 + \Delta\mu) 
  S_{\mu}^{\dag} S^{\mu} 
\nonumber \\
  &&
 +(\lambda_{\rm os} + a \gamma_{\rm os})
  \frac{\tilde{M}_8}{\sqrt{2}} \sum_{a} \delta_{a8}
  (\rho_{\mu}^{a \dag} S^{\mu}+\rho_{\mu}^{a} S^{\mu \dag})
\nonumber \\
  &&
 +(N \tilde{M}_0 \lambda_{\rm o1} + a^2 \beta_{\rm o} + N \tilde{M}_0 a \gamma_{\rm o1})
  \sum_{a}
  \rho_{\mu}^{a \dag} \rho^{\mu a}
\nonumber \\
  &&
 +(\lambda_{\rm o2} + a \gamma_{\rm o2})
  \sum_{a,b}
  (2 \tilde{M}_0 \delta_{ab} + \tilde{M}_8 d_{ab8})
  \rho_{\mu}^{a \dag} \rho^{\mu b},
\ea
where
\ba
  \beta_{\rm s}
  &=&
  N \beta_1 + N^2 \beta_2,\\
%\mbox{\hspace{10mm}}
  \beta_{\rm o}
  & = &
  2 \beta_5 + \beta_6 + 2 N \beta_7 + N \beta_8 + N^2 \beta_9,
\\
  \gamma_{\rm s}
  &=&
  \gamma_{1} + N \gamma_{2},\\
%\mbox{\hspace{10mm}}
  \gamma_{\rm os}
&  =&
  2 \gamma_{4} + N \gamma_{5},
\\
  \gamma_{\rm o1}
  &=&
  2 \gamma_{6} + \gamma_{7} + N \gamma_{8},\\
%\mbox{\hspace{10mm}}
  \gamma_{\rm o2}
&  =&
  2 \gamma_{9} + \gamma_{10} + N \gamma_{11}.
\ea
%
%===================
\subsection{Propagators}
\label{appendix:Prop}
%===================
%
The propagator for the singlet and octet vector meson is  given by
\bea
  G^{\rm S}_{\mu \nu}(r)
&  = &
  \frac{-i P_{\mu \nu}}{v \cdot r - a \alpha_{\rm s}},\\
%\mbox{\hspace{10mm}}
  G_{\mu \nu}^{{\rm O},ab}(r)
  & = &
  \frac{-i P_{\mu \nu} \delta^{ab}}{v \cdot r - a \alpha_{\rm o1}},
\eea
where $r_{\mu}$ is the residual momentum.

For the pseudo scalar meson, the propagator is
\be
  G^{ab}_{\rm PS}(p)
  =
  \frac{i \delta^{ab}}{p^2 - \tilde{m}_a^2},
\ee
where the 
leading order pseudo scalar mass $\tilde{m}_a^2$ of
Ref.\ \cite{Aoki:20051} is used.
The explicit form of the mass is
\be
  \tilde{m}^2_a
  =
  m_a^2 (1 - N c_3 - \tilde{c}_{3})
 -m_{\rm av}^2 N c_3 
 - N c_2 - \tilde{c}_{2},
\ee
with
\be
  m_{\rm av}^2
  =
  \frac{1}{N^2 - 1}
  \sum_a m_a^2,
\ee
where $m_a^2$ is the meson mass in the continuum limit
(i.e. $m_1^2=m_2^2=m_3^2=m_{\pi}^2$, $m_4^2=m_5^2=m_6^2=m_7^2=m_{K}^2$,
$m_8^2=m_{\eta}^2$).
$c_1$, $c_3$, $\tilde{c}_{3}$ are O($a$) terms \cite{Rupak:2002sm},
and $c_2$, $\tilde{c}_{2}$ are O($a^2$) terms \cite{Bar:2003mh,Aoki:2003yv}.
They are  low energy constants in the pure
pseudo scalar meson sector.
The mass $\tilde{m}_a^2$
is expressed in terms of the shifted quark masses,
\beqa
\tilde m_a^2 &=& \left\{
\begin{array}{ll}
\tilde m_\pi^2 = 2 B \tilde m, & a=1,2,3, \\
\tilde m_K^2 = B (\tilde m+\tilde m_s), & a=4,5,6,7,\\
\tilde m_\eta^2 = \dfrac{2B}{3}(\tilde m+2\tilde m_s), & a=8.\\
\end{array}
\right.
\eeqa

%\newpage
\subsection{Vertices}

\begin{itemize} 
\item NLO vector-vector 2-point vertex from 
${\cal L}^{0\pi}_{m}+{\cal L}^{0\pi}_{a^2}+{\cal L}^{0\pi}_{am}$

%See Fig. \ref{fig:OO},
%The singlet-singlet vertex is

%OO
 %%%%%%%%%%%%%%%%%%%%
 \begin{figure}[h]
  \begin{center}
      \psfragscanon
      \psfrag{O1}[][][2.5]{$\nu$, $b$}
      \psfrag{O2}[][][2.5]{$\mu$, $a$}
      \psfrag{text1}[4][2][2.5]{$
= i P_{\mu\nu} [ \delta_{ab} 
(N \tilde{M}_0 \lambda_{\rm o1} + a^2 \beta_{\rm o} + N a \tilde{M}_0 \gamma_{\rm o1})$}
      \psfrag{text2}[4][2][2.5]{$
+ (2 \tilde{M}_0 \delta_{ab} + \tilde{M}_8 d_{ab8})
(\lambda_{\rm o2} + a \gamma_{\rm o2})]$,}
      \scalebox{0.4}{\includegraphics{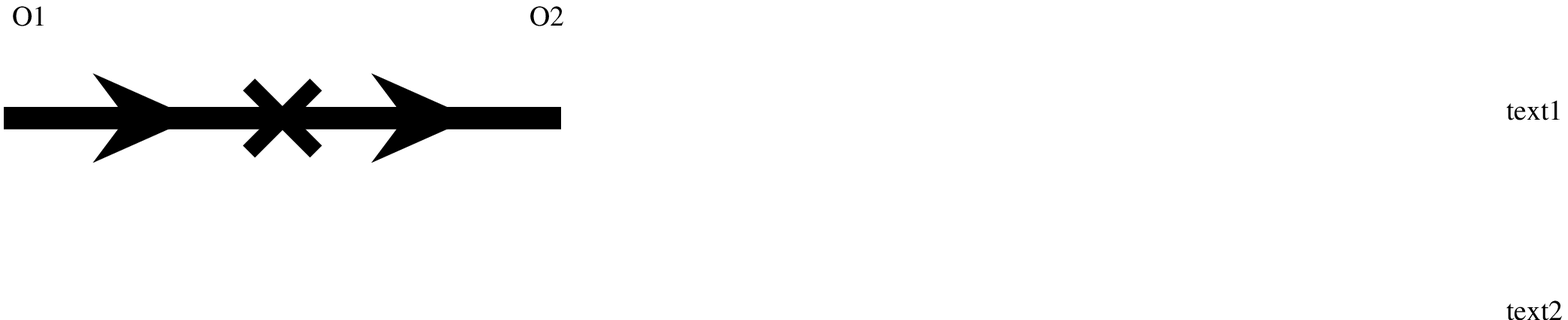}}
      \label{fig:OO}
  \end{center}
 \end{figure}
 %%%%%%%%%%%%%%%%%%%

%OS
 %%%%%%%%%%%%%%%%%%%%
 \begin{figure}[h]
  \begin{center}
      \psfragscanon
      \psfrag{O!}[][][2.5]{$\nu$, $a$}
      \psfrag{O2}[][][2.5]{$\mu$}
      \psfrag{text}[4][2][2.5]{$
= i P_{\mu\nu} \delta_{a8} \frac{\tilde{M}_8}{\sqrt{2}}
(\lambda_{\rm os} + a \gamma_{\rm os})$,}
      \scalebox{0.4}{\includegraphics{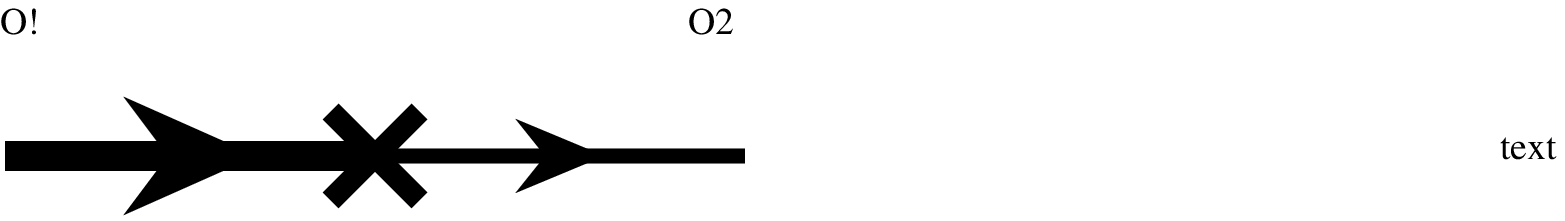}}
  \end{center}
 \end{figure}
 %%%%%%%%%%%%%%%%%%%

%SS
 %%%%%%%%%%%%%%%%%%%%
 \begin{figure}[h]
  \begin{center}
      \psfragscanon
      \psfrag{O1}[][][2.5]{$\nu$}
      \psfrag{O2}[][][2.5]{$\mu$}
      \psfrag{text}[4][2][2.5]{$
= i P_{\mu\nu} (N \tilde{M}_0 \lambda_{\rm s} + a^2 \beta_{\rm s} + N a
   \tilde{M}_0 \gamma_{\rm s} + \Delta\mu$).}
      \scalebox{0.4}{\includegraphics{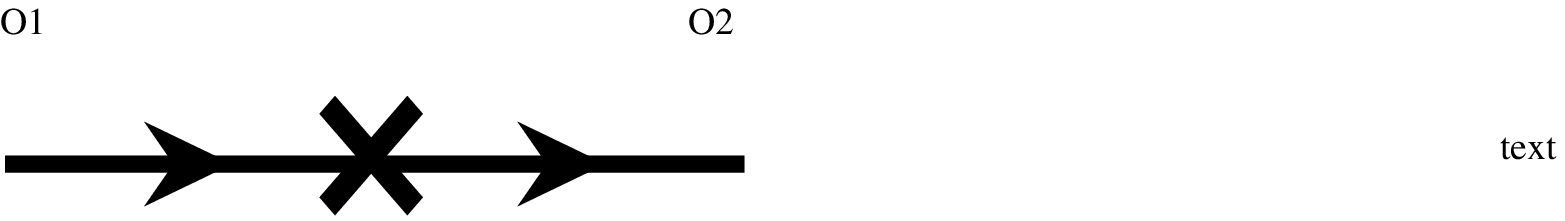}}
  \end{center}
 \end{figure}
 %%%%%%%%%%%%%%%%%%%

The oriented narrow line represents the singlet vector meson
and the oriented wide line represents the octet vector meson.

\newpage
\item LO vector-vector-pseudo-scalar 3-point vertex from
${\cal L}^{1\pi}_{p}$

%For single-octet-pseudoscalar vertex,

%OPS
 %%%%%%%%%%%%%%%%%%%%
 \begin{figure}[h]
  \begin{center}
      \psfragscanon
      \psfrag{P}[][][2.5]{$p$, $b$}
      \psfrag{O1}[][][2.5]{$\nu$, $a$}
      \psfrag{O2}[][][2.5]{$\mu$}
      \psfrag{text}[][][2.5]{$
=\frac{i g_1}{f} p_{\lambda} v_{\sigma}
\epsilon^{\mu \nu \lambda \sigma} \delta_{ab}$,}
      \scalebox{0.4}{\includegraphics{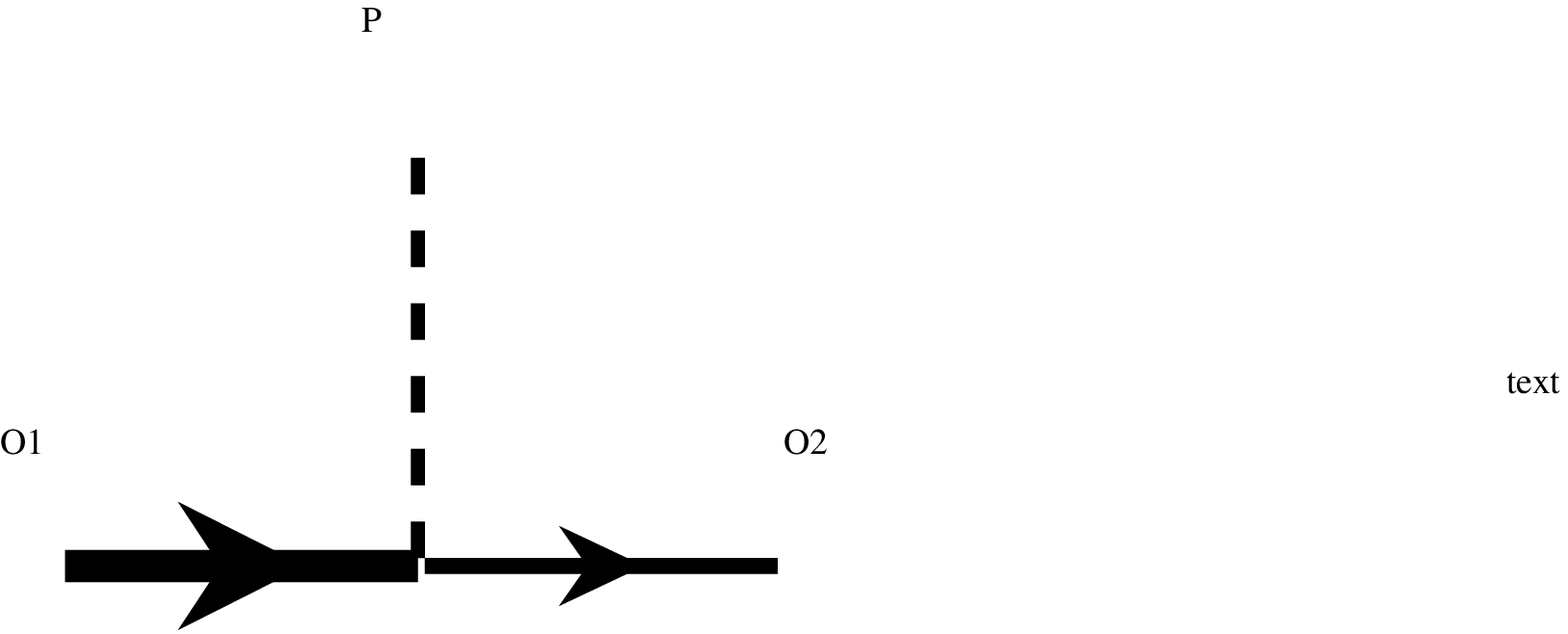}}
  \end{center}
 \end{figure}
 %%%%%%%%%%%%%%%%%%%
\vspace{-5mm}
%SPO
 %%%%%%%%%%%%%%%%%%%%
 \begin{figure}[h]
  \begin{center}
      \psfragscanon
      \psfrag{P}[][][2.5]{$p$, $b$}
      \psfrag{O1}[][][2.5]{$\nu$}
      \psfrag{O2}[][][2.5]{$\mu$, $a$}
      \psfrag{text}[][][2.5]{$
=\frac{i g_1}{f} p_{\lambda} v_{\sigma}
\epsilon^{\mu \nu \lambda \sigma} \delta_{ab}$,}
      \scalebox{0.4}{\includegraphics{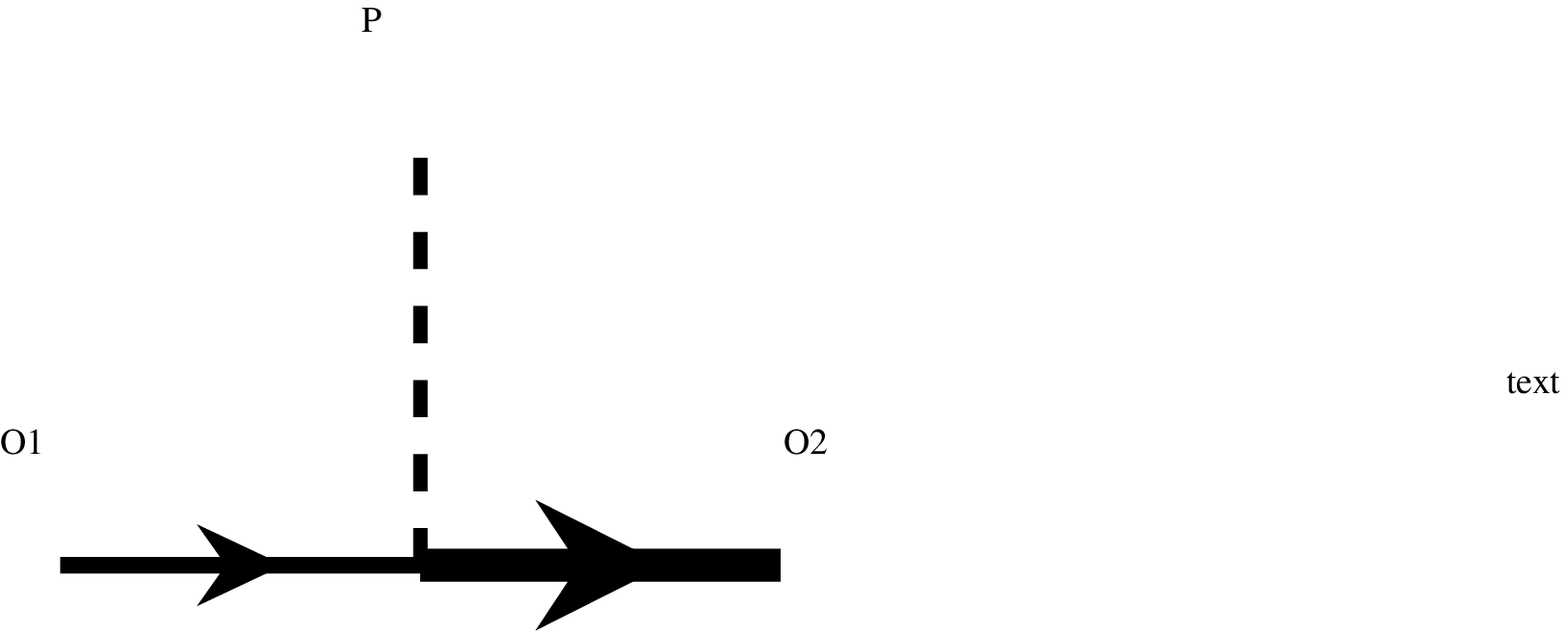}}
  \end{center}
 \end{figure}
 %%%%%%%%%%%%%%%%%%%
\vspace{-5mm}
%OPO
 %%%%%%%%%%%%%%%%%%%%
 \begin{figure}[h]
  \begin{center}
      \psfragscanon
      \psfrag{P}[][][2.5]{$p$, $c$}
      \psfrag{O1}[][][2.5]{$\nu$, $b$}
      \psfrag{O2}[][][2.5]{$\mu$, $a$}
      \psfrag{text}[][][2.5]{$
=\frac{i g_2 \sqrt{2}}{f} p_{\lambda} v_{\sigma}
\epsilon^{\mu \nu \lambda \sigma} d_{abc}$.}
      \scalebox{0.4}{\includegraphics{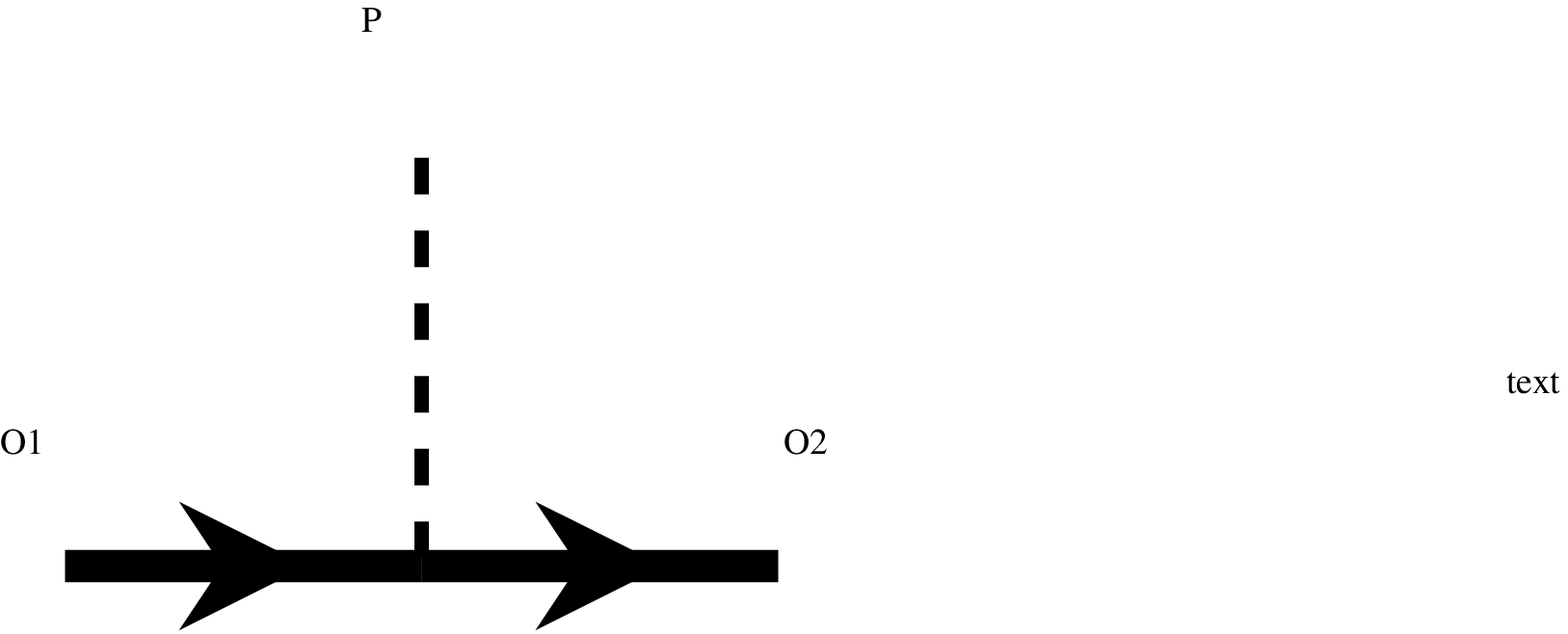}}
  \end{center}
 \end{figure}
 %%%%%%%%%%%%%%%%%%%
The dotted line represents the incoming
pseudo scalar meson.

\item LO vector-vector-pseudo scalar-pseudo scalar 4-point vertex
from ${\cal L}^{2\pi}_{a}$

 %%%%%%%%%%%%%%%%%%%%
 \begin{figure}[h!]
  \begin{center}
   \begin{tabular}{c}
%OPPO vertex
      \psfragscanon
      \psfrag{P1}[][][2.5]{$d$} 
      \psfrag{P2}[][][2.5]{$c$}
      \psfrag{O1}[][][2.5]{$\nu$, $b$}
      \psfrag{O2}[][][2.5]{$\mu$, $a$}
      \psfrag{text}[][][2.5]{$
= i P_{\mu\nu} [ - a \frac{4}{N f^2}
( \alpha_{\rm o1} \delta_{ab} \delta_{cd}
+ \alpha_{\rm o2} \sum_{e} d_{abe} d_{cde})]$,}
      \scalebox{0.4}{\includegraphics{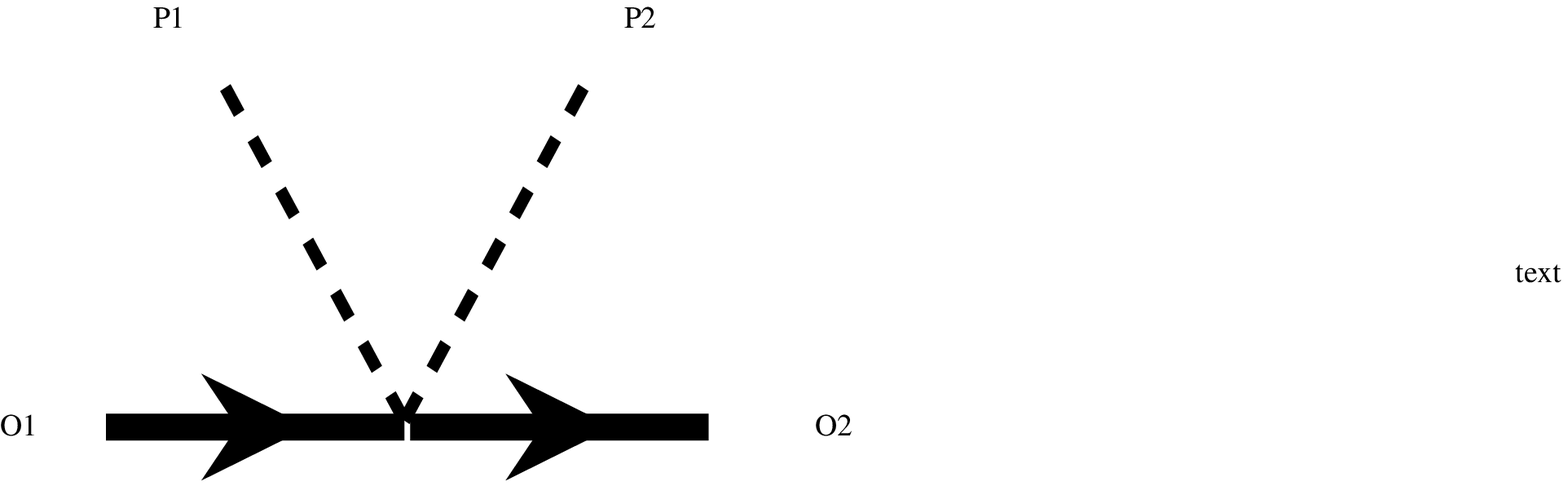}}
\vspace{6mm}
\\
%OPPS vertex
      \psfragscanon
      \psfrag{P1}[][][2.5]{$b$}
      \psfrag{P2}[][][2.5]{$c$}
      \psfrag{O1}[][][2.5]{$\nu$, $a$}
      \psfrag{O2}[][][2.5]{$\mu$}
      \psfrag{text}[][][2.5]{$
= i P_{\mu\nu} [ - a \frac{2\sqrt{2}}{N f^2} \alpha_{\rm os} d_{abc}]$,}
      \scalebox{0.4}{\includegraphics{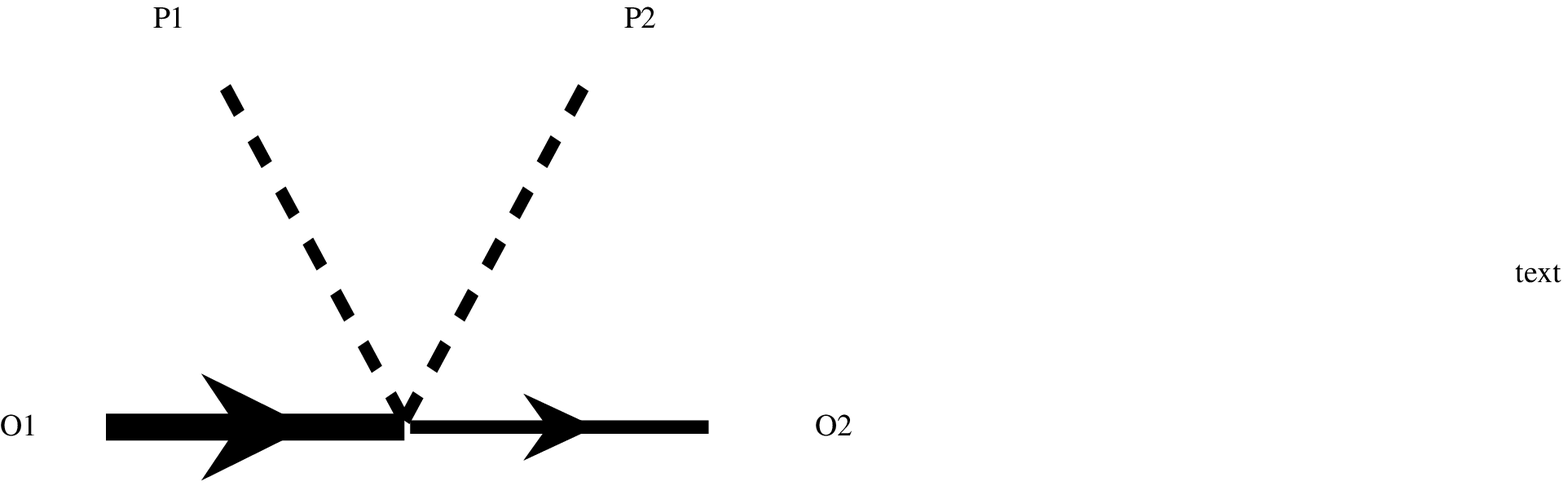}}
\vspace{6mm}
\\
%SPPS vertex
      \psfragscanon
      \psfrag{P1}[][][2.5]{$a$}
      \psfrag{P2}[][][2.5]{$b$}
      \psfrag{O1}[][][2.5]{$\nu$}
      \psfrag{O2}[][][2.5]{$\mu$}
      \psfrag{text}[][][2.5]{$
= i P_{\mu\nu} [ - a \frac{4}{N f^2} \alpha_{\rm s} \delta_{ab}]$.}
      \scalebox{0.4}{\includegraphics{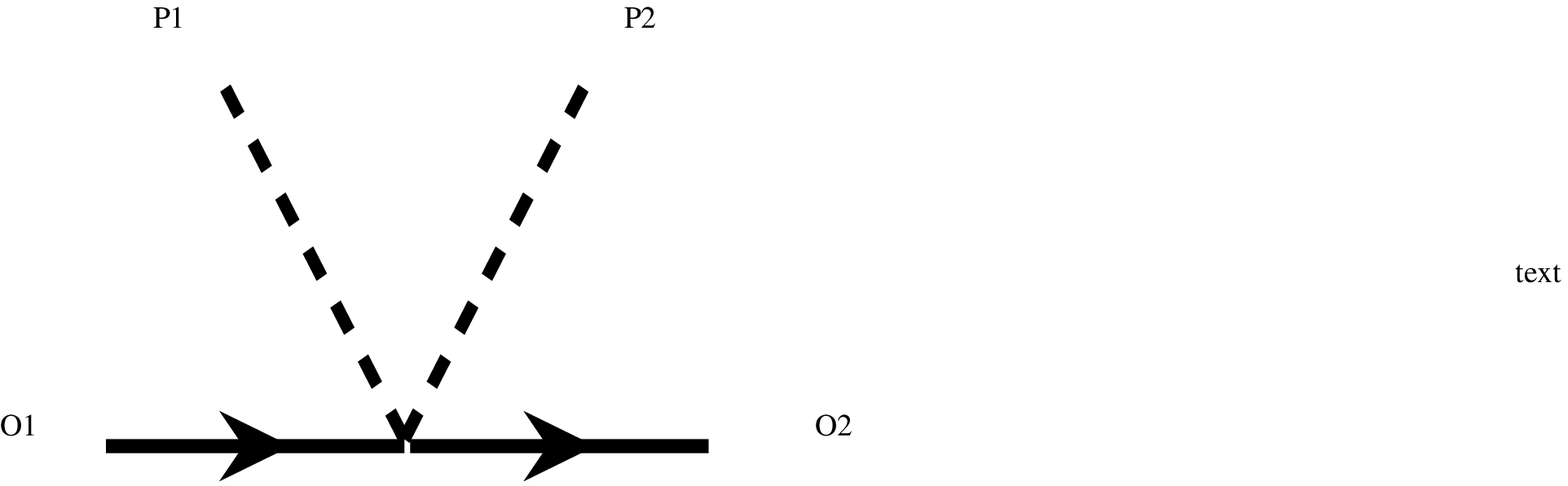}}
   \end{tabular}
  \end{center}
 \end{figure}
 %%%%%%%%%%%%%%%%%%%
%Note that we do not show ${\cal L}^{2\pi}_{p}$ here.
\end{itemize}

%\newpage
%
%%%%%%%%%%%%% Bibliography
%\bibliographystyle{physics_tst}
%\bibliography{biblioVChPT}

\end{document}